\documentclass[apj]{emulateapj}

\shorttitle{32-band Photometric Redshifts in the ECDF-S}
\shortauthors{Cardamone et al.}
\begin{document}
\title{The Multiwavelength Survey by Yale-Chile (MUSYC): Deep Medium-Band optical imaging and high quality 32-band photometric redshifts in the ECDF-S\footnote{\sc Based [in part] on data collected at  Subaru Telescope, which is operated by the National Astronomical Observatory of Japan.}}
\author{Carolin N. Cardamone\altaffilmark{1,2},
Pieter G.\ van Dokkum \altaffilmark{1},  
C.\ Megan Urry\altaffilmark{1,2},
Yoshi Taniguchi\altaffilmark{3}, Eric Gawiser\altaffilmark{4}, Gabriel Brammer\altaffilmark{1}, Edward Taylor\altaffilmark{5,6}, Maaike Damen\altaffilmark{5}, Ezequiel Treister\altaffilmark{7} ,  Bethany E. Cobb\altaffilmark{8}, Nicholas Bond\altaffilmark{4}, Kevin Schawinski\altaffilmark{2,9}, Paulina Lira\altaffilmark{10}, Takashi Murayama \altaffilmark{11}, Tomoki Saito \altaffilmark{13},Kentaro Sumikawa \altaffilmark{12}}


\email{carolin.cardamone@yale.edu}
\altaffiltext{1}{Department of Astronomy, Yale University, New Haven, CT 06511, USA}
\altaffiltext{2}{Yale Center for Astronomy and Astrophysics, Yale University, P.O.~Box 208121, New Haven, CT 06520}
\altaffiltext{3}{Research Center for Space and Cosmic Evolution, Ehime University, Bunkyo-cho 2-5, Matsuyama 790-8577, Japan}
\altaffiltext{4}{Department of Physics \& Astronomy,Rutgers University, Piscataway, NJ 08854-8019}
\altaffiltext{5}{Sterrewacht Leiden, Leiden University, NL-2300 RA Leiden, Netherlands}
\altaffiltext{6}{School of Physics, the University of Melbourne, Parkville, 3010, Australia}
\altaffiltext{7}{Institute for Astronomy, 2680 Woodlawn Drive, University of Hawaii, Honolulu, HI 96822, U.S.A.}
\altaffiltext{8}{Department of Astronomy, University of California, Berkeley, CA 94720-3411}
\altaffiltext{9}{Einstein Fellow}
\altaffiltext{10}{Departamento de Astronoma, Universidad de Chile, Casilla 36-D, Santiago, Chile}
\altaffiltext{11}{Astronomical Institute, Graduate School of Science, Tohoku University, Aramaki, Aoba, Sendai 980-8578, Japan}
\altaffiltext{12}{Graduate School of Science and Engineering, Ehime University,  Bunkyo-cho, Matsuyama 790-8577, Japan}
\altaffiltext{13}{Institute for Physics and Mathematics of the Universe, The University of Tokyo,5-1-5 Kashiwanoha, Kashiwa, Chiba 277-8583, Japan}

\begin{abstract}
We present deep optical 18-medium-band photometry from the Subaru
telescope over the $\sim 30'\times 30'$ Extended Chandra Deep Field-South (ECDF-S), as part
of the Multiwavelength Survey by Yale-Chile (MUSYC). This field has
a wealth of ground- and space-based ancillary data, and contains
the GOODS-South field and the Hubble Ultra Deep Field.
We combine the Subaru imaging with existing $UBVRIzJHK$ and {\it Spitzer}
IRAC images to create a uniform catalog.
Detecting sources in the MUSYC ``$BVR$''
image we find $\sim$ 40,000 galaxies with $R_{AB}<25.3$, the median
5$\sigma$ limit of the 18 medium bands. 
Photometric redshifts  are determined using the EAZY code and compared to $\sim 2000$
spectroscopic redshifts in this field.
The medium band filters provide very accurate redshifts for the
(bright) subset of galaxies with spectroscopic redshifts, particularly
at $0.1<z<1.2$ and at $z\ga3.5$. For $0.1<z<1.2$, we find a $1\sigma$ scatter
in $\Delta z/(1+z)$ of $0.007$, similar to results obtained with
a similar filter set in the COSMOS field. 
As a demonstration of the data quality, we show that the red sequence and blue cloud can be
cleanly identified in rest-frame color-magnitude diagrams
at $0.1<z<1.2$.
We find that $\sim$20\% of the red-sequence-galaxies show evidence of dust-emission at longer rest-frame wavelengths.  
The reduced images, photometric catalog, and photometric redshifts are
provided through the public MUSYC website.
\end{abstract}

\keywords{cosmology: observations ---
galaxies: evolution --- galaxies:
formation --- astronomical databases: surveys
 --- astronomical databases: catalogs
}

\section{INTRODUCTION}
\label{intro}
Studies of distant galaxies and active galactic nuclei (AGN) require
redshift measurements to determine distances and associated lookback times.
For optically-bright objects, redshifts can be measured efficiently with
multi-slit spectrographs on large telescopes \citep{davisetal2003,lefevreetal2005}.
However, spectroscopic redshift measurements are very difficult
for galaxies and AGN that are at $z>1$,
obscured, or intrinsically faint. As a result, we had to rely
almost exclusively on photometric redshifts determined from broad-band
photometry for galaxies with $I\gtrsim 23$ \citep[see, e.g., ][, and many other studies]{grazianetal2006, wuytsetal2008}.

The COMBO-17 survey \citep{wolfetal2004}
pioneered the use of medium-bandwidth filters as
a compromise between imaging and spectroscopy. These filters sample the
spectral energy distributions (SEDs) of galaxies at a resolution of
$R=10-20$ and provide a redshift quality of 1--2\,\%,
intermediate between spectroscopy and broad-band imaging. This
improvement enables measurements of rest-frame
colors and  the environment of galaxies, and improves the accuracy
of determinations of the physical properties of galaxies. It also
opens up the possibility to directly detect strong emission lines, which is
particularly relevant for the identification of AGN.

With the development of medium-band filter sets in the near-IR
\citep{vandokkumetal2009newfirm} and on 8m class
telescopes \citep{Ilbertetal2009}, accurate redshifts are now becoming
available for objects that are well beyond the limits of spectroscopy,
and the full potential of this technique is being realized.

In this paper, we present deep 18-band optical medium-band photometry
from the Subaru telescope in the $\sim 30'\times 30'$
``Extended'' Chandra Deep Field-South, as part of the Multiwavelength
Survey by Yale-Chile (MUSYC; \citealt{gawiseretal2006}).
The ECDF-S field has an extensive set of ancillary data
\citep[see ][]{tayloretal2009}, recently augmented by very
deep {\it Spitzer} IRAC data in the
SIMPLE\footnote[3]{\url{http://www.astro.yale.edu/dokkum/SIMPLE/}}
survey (Damen et al. 2010, submitted). Furthermore, the field contains
the GOODS-South field and the Hubble Ultra Deep Field.
Our medium band survey is similar to that of Ilbert et al.\ (2009) in
the larger COSMOS field, although we
include 18 medium bands while there are only 12 available
for the larger COSMOS field. 

We discuss the observations in Section \ref{sec:obs} and the details of data reduction in Section \ref{sec:data}. 
We include ancillary data in our catalog from the MUSYC survey, which is described in Section \ref{sec:adddata}. 
The details of our photometry are discussed and the public catalog is presented in Section \ref{sec:phot}.
We describe the photometric redshift determinations and evaluate the improvements provided by medium-band filters in Section \ref{sec:redshifts}.
Finally in Section \ref{sec:redseq} we present the  color-magnitude
diagrams of the ECDF-S out to redshift 1.2 and investigate the occurrence of dusty galaxies on the red sequence.
Throughout this paper we assume $H_0 = 71  \: {\rm km \: s^{-1} \: Mpc^{-1}}$, $\Omega_{\rm m} = 0.3$ and $\Omega_{\rm \Lambda} = 0.7$.

\section{Observations}
\label{sec:obs}
\begin{figure}
\includegraphics[angle=0, width=0.47\textwidth]{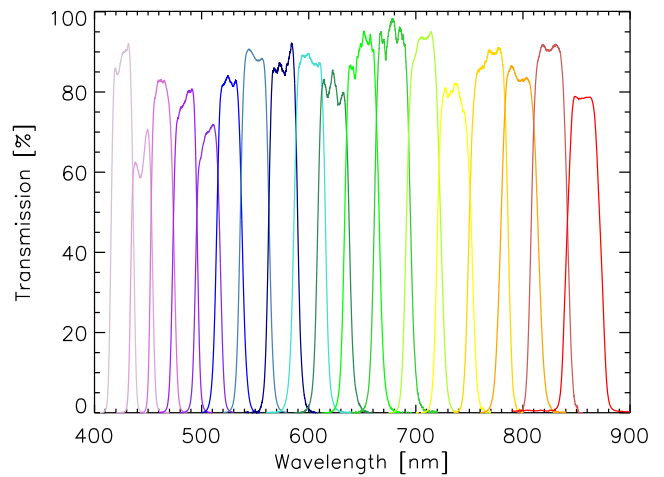}
\caption{Subaru medium-band filter transmission curves, including atmospheric transmission, quantum efficiency, and the transmission of the optical elements of the telescope plus instrument.  This filter set was designed to sample the spectrum evenly between 400 and 900 nm with an effective resolution of $\lambda / \Delta \lambda \sim23$ in order to compute accurate photometric redshifts.}
\label{response}
\end{figure}

We carried out a medium-band imaging campaign on the Subaru Telescope using the wide field-of-view camera Suprime-Cam \citep{miyazakietal2002,iyeetal2004}.
Observations were collected in 6 runs over 2 years, in January, February and December 2006, and January, March and December 2007; the observations are summarized in Table \ref{tab:obs}. 
We show the response curves of the medium-band filters used in this study in Figure \ref{response}.  These filters were designed to maximize the accuracy of photometric redshifts by evenly sampling the wavelength range from 400-900 nm with an effective resolution of $\lambda / \Delta \lambda \sim23$ \citep{taniguchi2004}. 

\begin{figure}
\includegraphics[angle=0, width=0.47\textwidth]{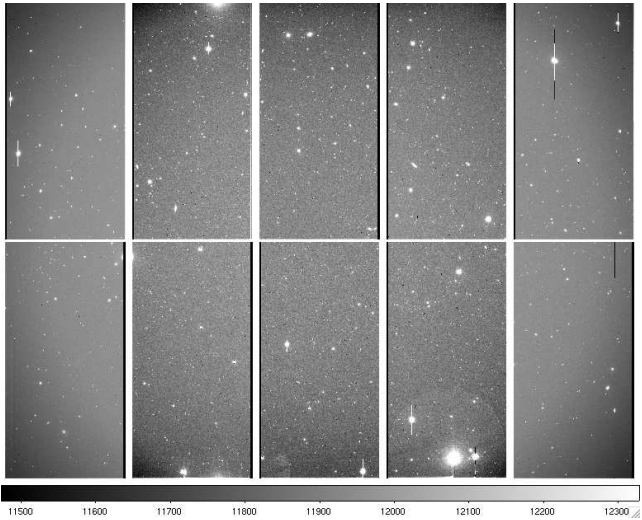}
\caption{Example raw data frame from Suprime-Cam, for a 600s exposure in Filter IA651 taken on 2007-1-18 (East-up, North-right).  Each exposure is composed of 10 CCDs in a 5 x 2 array with small gaps between each chip.  There are large gradients in the background illumination of each chip from sensitivity variations and vignetting.  There are bad pixel columns and regions of low sensitivity near the chip edges, which must be accounted for in the data reduction.}
\label{raw}
\end{figure}

Suprime-Cam consists of ten 2k $\times $ 4k CCD chips that cover a large field of view (24\arcmin  $\times$ 27\arcmin).
A single raw exposure in filter IA651 is shown in Figure \ref{raw}; each of the 10 individual chips show large gradients in background (from sensitivity variations and vignetting), as well as regions of bad pixels, and there are small gaps between the individual CCDs.
Therefore, our observing strategy consisted of taking multiple Suprime-Cam exposures, dithering the pointing slightly to cover the small chip gaps.
The median number of individual exposures for a given filter was 9, but the number of exposures and the exposure time were varied depending on conditions at the telescope and the available time each night.
The long exposure on $IA~ 738$ was designed to be used as a detection image  for
relevant science projects, but for our purposes here we prefer the significantly deeper MUSYC $BVR$ image as a detection image.
Table \ref{tab:obs} lists each filter (1), the full-width at half maximum in nanometers (2), its observation date (3), the number of dithered exposures (4), and the total exposure time in seconds (5).
\begin{table}
 \caption{Medium Band Observations}
 \label{tab:obs}
\begin{center}
 \begin{tabular}{@{}cccrr}
 \hline
 \hline
{Band} &  {FWHM [nm]} & {Date Observed}  & {\# Exp} & {Exptime [s]} \\
{(1)} & {(2)} & {(3)} &{(4)} &{(5)} \\
 \hline
IA427  & 21 & 2006-01-30 &   7 &  2100   \\
IA445  & 20 & 2007-01-19 &   9 &  5400   \\
IA464  & 22 & 2006-01-31 &   7 & 2940   \\
IA484  & 23 & 2007-01-17 &   9 &  5400   \\
IA505  & 26 & 2006-01-30 &    8&  2400   \\
IA527  & 24 & 2007-01-17 &    9 & 5400   \\
IA550  & 28 & 2007-01-19 &    9 & 5400   \\
IA574  & 27 & 2006-01-30 &    7 & 2100   \\
IA598  & 30 & 2007-01-18 &    8 &4800   \\
IA624  & 30 & 2006-12-20 &   10 & 6000   \\
IA651  & 33 & 2007-01-18 &   9 &  5400   \\
IA679  & 34 & 2006-12-18 &   7 &  6300   \\
IA709  & 32 & 2006-01-29 &   7 &  4200   \\
IA738  & 33 & 2007-01-16 &   19 & 11400   \\
IA767  & 37 & 2006-12-19 &   5 &  4200   \\
IA797  & 35 & 2006-12-19 &   4 &  4800  \\
IA827  & 34 & 2006-01-28 &   4 &  3600   \\
IA856  & 34 & 2006-12-20 &   4 &  2400   \\
 \hline
\end{tabular}
\end{center}
\end{table}

\section{Data Reduction}
\label{sec:data}
The data were reduced using a combination of standard \texttt{IRAF}\footnote{IRAF is distributed by the National Optical Astronomy Observatory, which is operated by the Association of Universities for Research in Astronomy, Inc., under a cooperative agreement with the National Science Foundation.} tasks; \texttt{sdfred}, a data reduction and analysis software package written for Suprime Cam \citep{yagietal2002, ouchietal2004}; and custom tasks.   Our reduction procedures follow those of \citet{labbeetal2003} and \citet{quadrietal2007}, and are briefly described here.

\subsection{Flat Fielding and Bias Subtraction}
Dome flats are well-exposed with high Signal-to-Noise ratios (S/N) and show the pixel-to-pixel variations on the CCD, while dark sky flats more accurately reproduce the variation in sensitivity to the night sky spectrum and the illumination pattern across the CCDs.
Therefore, for each filter both dome flats and dark sky flats are constructed.
We divide the high S/N dome flats by the dark sky flats, effectively removing the differences between the CCD response to the dome lights and the light of the night sky, i.e., flattening the dome flats.
To maintain the high S/N, we then smooth the resulting flats with the \texttt{iraf} routine \texttt{boxcar}.  
The smoothing kernel for the \texttt{boxcar} routine was selected to be 10 pixels on a side because this minimizes the variance between adjacent smoothed areas ($\le$0.1\%); this effectively increases our S/N by an order of magnitude.

We also constructed master bias frames, combining $\sim$10-20 bias frames collected each observing run with the \texttt{IRAF} routine \texttt{zerocombine}. One master bias frame per run was sufficient as the bias frames were stable from night to night.

After we construct the flat and bias frames, we apply them to the raw data frames.
First, the two-dimensional master bias frames are subtracted from each raw data frame, then the remaining overscan correction is subtracted using \texttt{sdfred}.
Next, flat fielding is done by first dividing the raw science exposures by the dome flat and then dividing this by our smoothed sky flat, as follows:
\begin{equation}
\frac{rawframe-bias}{domeflat}\times(\frac{domeflat}{skyflat})_{smoothed}
\end{equation}
Even after careful flat fielding, small gradients in the sky background can remain that can affect photometry non-uniformly across the field.
Using \texttt{SExtractor} \citep{bertinarnouts1996}, we measured and removed the sky gradient across the field using a large 2D mesh (1000 pixels on a side).
The large mesh was selected to avoid overestimating the sky near the edges of extended objects and near the location of faint objects not detected in individual exposures.

\subsection{Image Combination}
\label{sec:comb}
After flat fielding and bias subtraction, we are left with multiple individual exposures for each filter, each composed of 10 chips, ready to be combined.
These chips must be aligned and then combined, preserving the flux in the stars and rejecting bad pixel artifacts.

Suprime-Cam provides a very large field-of-view, but with large geometric distortions caused by the optics.  
These distortions must be corrected for when combining individual dithered data frames into a final image.
For a first-order coordinate correction, we use the standard task in \texttt{sdfrd}.
Then we apply a secondary correction using the deep $BVR$ combined image from the MUSYC Survey \citep{gawiseretal2006} using the \texttt{iraf} routine \texttt{mscimatch}\footnote{The MSCRED package in IRAF was originally developed for mosaic reductions by NOAO. Information about the NOAO Mosaic Project can be found at \url{http://www.noao.edu/kpno/mosaic/mosaic.html}.}.

In our image combination, we maximize the S/N in the seeing disk of point sources by combining the individual CCD chips from each exposure using a weighted average.
The weighted average is determined by the RMS of the sky, the average FWHM and the relative flux scale for each of the individual CCD exposures, following the procedure described in detail in Appendix A of \citet{gawiseretal2006}, which was used in creating the broad-band point-source-optimized images for the MUSYC survey.
This results in an optimal weight for each individual CCD chip as they are combined to make the final image.
The weights are normalized to the first exposure central bottom chip.

Additionally, to improve our image quality, we use carefully constructed bad pixel maps when combining the final images.
Bad pixel maps are first created from pixels marked by \texttt{sdfred}, including bad columns, isolated dead pixels and the CCD chip edges.  
Then careful visual inspection is made of each exposure and all bad pixels found by eye, in addition to artifacts such as satellite trails, are included in the final bad pixel maps.

\begin{figure*}
\includegraphics[angle=0, width=0.99\textwidth]{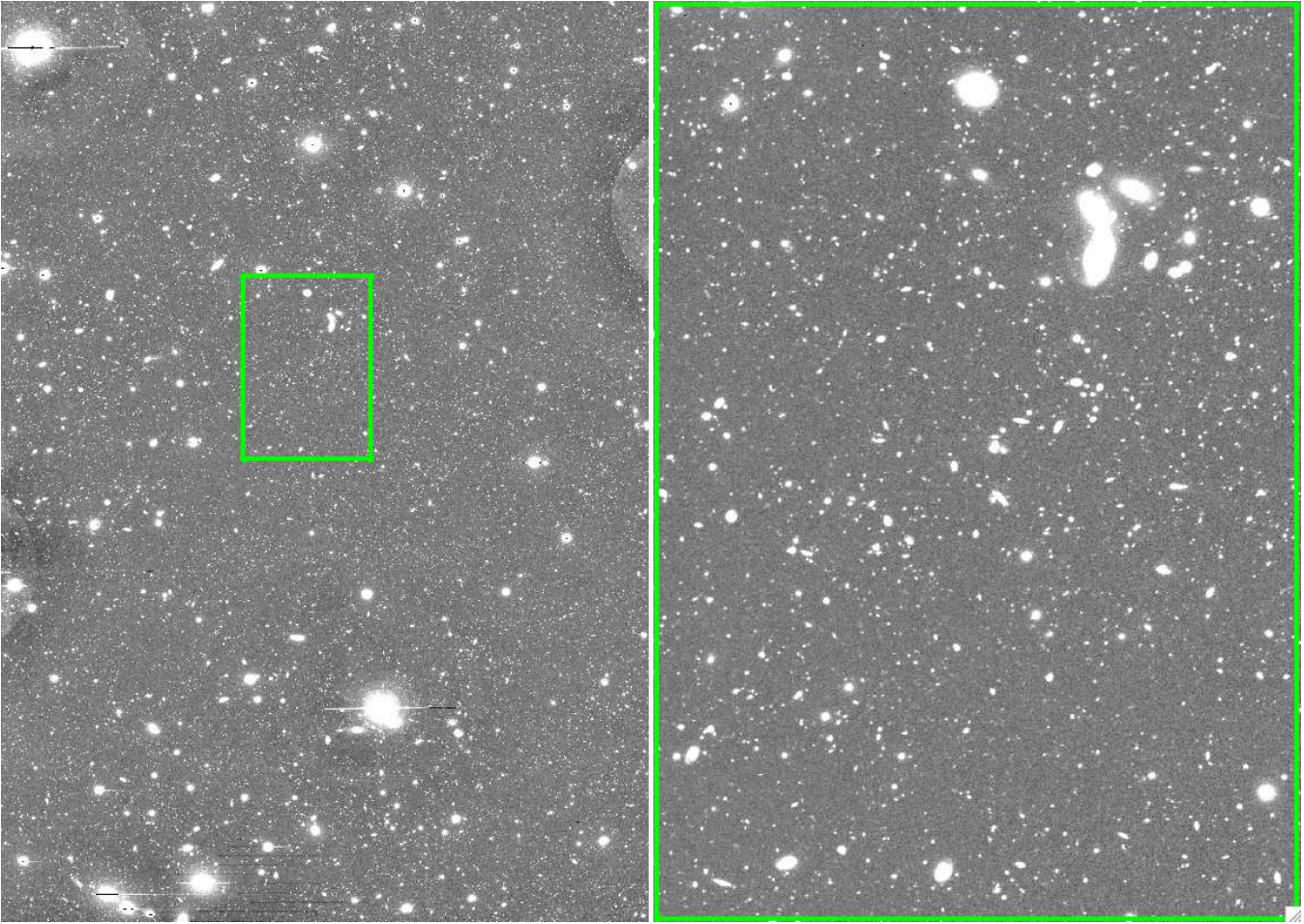}
\caption{{\it Left:} Full frame showing the final combined image in filter $IA~ 651$, rotated 90 degrees from the raw frame shown in Figure \ref{raw} (North is up, East is to the left).  Note the uniformity across the field from careful attention to flat fielding and the dithering pattern of individual exposures.  The S/N at the edge of the field is lower due to the smaller number of individual exposures combined at the edges with our dither pattern.  A box indicates the region of the field displayed in the right panel, zooming in by a factor of 5.  {\it Right:} Detail of a final combined image in the filter IA651.  Notice the detailed morphology visible for nearby large galaxies. }
\label{fig:final}
\end{figure*}

The final images are then created by averaging each of the individual weighted chips, rejecting the bad pixels and using the \texttt{iraf} routine \texttt{combine} with a percentile clipping algorithm.  
Experimentation showed the percentile clipping algorithm provided optimal cosmic ray rejection.
In Figure \ref{fig:final}, we show final reduced images for $IA~ 651$.
On the left of Figure \ref{fig:final}, we show the entire frame to show the flatness we achieved across the image, while the {\it right} panel zooms in on region displayed in the green box in the {\it left} panel.
In the zoom-in, we see that many galaxies are clearly visible to the eye, the local background is relatively flat and no cosmic rays remain.
We list details about each final combined image in Table \ref{tab:subaru}.  
The FWHM, Column 2, is calculated using the \texttt{iraf} routine \texttt{imexam} on several hundred stars.

\subsection{Photometric Calibration}
\label{sec:zpt}
We now have  a single final image for each filter, but these must be placed on a standard flux scale.
To achieve this, each night at the telescope a handful of ESO spectro-photometric standard stars\footnote{http://www.eso.org/sci/observing/tools/standards/spectra/} were observed.
These stars are used to calculate the zero-point for each final image.

We calculate the magnitudes for these standard stars in our filter system ($m_{ss}$) by convolving the ESO standard star spectra with the effective transmission in each filter.  
We can then calculate the zero-point for each filter, using the amount of flux observed in each star ($flux_{obs}$) in the standard star images.  
The zero point for the final image in each filter is given by:\\
\begin{equation}
 m_{zp} = m_{ss} - 2.5 \times \log{flux_{obs}} - airmass \times k_\nu 
 \end{equation}
Here $k_\nu$, the air-mass coefficient, stands for the extinction in magnitudes per airmass and is necessary because each standard star observation was made at a different airmass from the science images.
To determine $k_\nu$, we first obtained the Extinction Curve for Mauna Kea from the CFHT observers' manual\footnote{http://www.cfht.hawaii.edu/Instruments/ObservatoryManual/CFHT\_Observatory\_Manual\_TOC.html}  and compared this to an extinction curve derived on Mauna Kea in the $B$ and $V$ bands published by \citet{krisciunasetal1987}.
Additionally, on nights when a single standard star was observed at  multiple airmasses, we directly calculated extinction curves.
All $k_\nu$ measurements were consistent with each other.
Therefore when calculating the zero points for each filter, we used our own measured values for the six filters for which they were available ($IA427, IA484, IA527, IA624, IA738$, and $IA856$), two additional values from \citet{krisciunasetal1987}; the $B$ band for $IA~ 445$ and the $V$ band for $IA~ 550$ and interpolated values for the remaining 10 filters according to the curve published by the CFHT.
The fluxes are normalized such that they are fluxes per second, to account for the different exposure times of the standard stars and the final image exposures.  
The air-mass assigned to the final combined image is that of the first exposure, which is used to normalize the flux in the image combination (\S\ref{sec:comb}).
The zero points for each filter are listed in Table \ref{tab:subaru}, Column 4.

\subsection{Noise Properties}
\label{sec:noise}
The noise properties of each final image must be understood in order to calculate the nominal depth of each final combined image, as well as accurate photometric errors.
SExtractor assumes Poisson sky noise and thus can underestimate the total errors in the photometry, which include contributions from electron read-out noise, sky noise, imperfect background subtraction and pixel-to-pixel correlations that can be introduced during the reduction process. 
To accurately account for all of these effects, we sum the counts in a large number of apertures randomly placed throughout the image, and add these error estimates in quadrature with the output error from SExtractor \citep{labbeetal2003,gawiseretal2006,quadrietal2007}.
The apertures are placed in locations to avoid objects (using SExtractor's segmentation map) and have an identical size to the photometry apertures we later use in each band (See Section \ref{sec:phot}).
The roughly Gaussian distribution of flux in these apertures describes the noise in the background on the image, and the sigma width of this Gaussian measures the uncertainty in the background noise (i.e., the depth) of the imaging. 
These depths are based on flux measurements within the aperture size used for photometry.
Integrating over the PSF, an offset of $-0.6$ magnitudes is necessary to correct these magnitudes from aperture values to total values (see \S\ref{sec:apsel}).
We report 5$\sigma$ depths with this correction in Table \ref{tab:subaru}, Column 3.
We note an independent analysis of these same data give slightly different zeropoints and depths, due to small differences in methodology.  The zeropoints are consistent to $\sim0.1$ mag for all filters.

\subsection{Final Images}
We created final combined images for each filter (Figure \ref{fig:final}), which are available from the MUSYC webpage\footnote{\texttt{http://physics.rutgers.edu/\~{}gawiser/MUSYC}}.
We list each image and its properties in Table \ref{tab:subaru}:  Column 1 lists the central wavelength of the filter in nm, Column 2 lists the seeing, as measured using the average FWHM of a sample of several hundred stars in the actual image, Column 3 lists the 5 $\sigma$ depth in AB magnitudes, and Column 4 lists the zero point value for the image.

\begin{table}
 \caption{Medium Band Image Properties}
 \label{tab:subaru}
\begin{center}
 \begin{tabular}{@{}cccc}
 \hline
 \hline
{Band} & {FWHM [\arcsec]} &{5$\sigma$ depth [AB]\tablenotemark{1} } & {Zero Point [AB]} \\
{(1)} & {(2)} & {(3)} & {(4)} \\
 \hline
IA427  &  1.01 &   25.01    &  25.10 $\pm$ 0.11      \\
IA445  &  1.23 &   25.18    &  25.07 $\pm$ 0.08      \\
IA464  &  1.79 &   24.38    &  25.30 $\pm$ 0.03      \\
IA484  &  0.76 &   26.22    &  25.50 $\pm$ 0.05      \\
IA505  &  0.94 &   25.29    &  25.34 $\pm$ 0.02      \\
IA527  &  0.83 &   26.18    &  25.72 $\pm$ 0.03      \\
IA550  &  1.13 &   25.45    &  25.88 $\pm$ 0.06      \\
IA574  &  0.95 &   25.16    &  25.71 $\pm$ 0.02      \\
IA598  &  0.63 &   26.05    &  26.02 $\pm$ 0.03      \\
IA624  &  0.61 &   25.91    &  25.89 $\pm$ 0.05      \\
IA651  &  0.60 &   26.14    &  26.15 $\pm$ 0.03      \\
IA679  &  0.80 &   26.02    &  26.20 $\pm$ 0.03      \\
IA709  &  1.60 &   24.52    &  26.02 $\pm$ 0.03      \\
IA738  &  0.77 &   25.93    &  26.02 $\pm$ 0.02      \\
IA767  &  0.70 &   24.92    &  26.04 $\pm$ 0.02      \\
IA797  &  0.68 &   24.69    &  26.02 $\pm$ 0.02      \\
IA827  &  1.69 &   23.60    &  25.92 $\pm$ 0.04      \\
IA856  &  0.67 &   24.41    &  25.73 $\pm$ 0.01      \\
 \hline
\vspace{-0.6cm}
\footnotetext[1]{Total Magnitude}
\end{tabular}
\end{center}
\end{table}

\section{Ancillary Data}
\label{sec:adddata}
\subsection{MUSYC Survey: Existing Optical and Near-Infrared Data}
The Extended {\it Chandra} Deep Field-South (ECDF-S) has been targeted by a host of optical and infrared surveys (\citealt{arnoutsetal2001,moyetal2003,wolfetal2004, gawiseretal2006,Hildebrandtetal2006}; Damen et al 2010, submitted).
We obtain reduced optical and near-infrared imaging from \citet{tayloretal2009}, where the observations, reductions and characteristics of these data are described in detail.
To summarize briefly, the $UU_{38}BVRI$ imaging originate from the ESO archive and were combined from multiple projects using the Wide Field Imager (WFI) on the ESO MPG 2.2 m telescope.  These data were collected and calibrated as part of GaBoDS (the Garching-Bonn Deep Survey;  \citealt{Hildebrandtetal2006}).  
The $z$-band data, collected as part of the MUSYC survey, are from the Mosaic-II camera on the CTIO 4m Blanco telescope \citep{mulleretal1998} and are described further by \citet{gawiseretal2006} and \citet{tayloretal2009}.
The $H$-band data, taken with SofI on the ESO NTT 3.6 m telescope \citep{moyetal2003}, covers 80\% of the field \citep{tayloretal2009}.
The $JK$ imaging was obtained using the ISPI camera on the CTIO Blanco 4m telescope \citep{tayloretal2009}.
The images along with the size of the FWHM, 5$\sigma$ depth, zero point and survey in which they were observed, are listed in Table \ref{tab:broad}.
The FWHM and 5$\sigma$ depth are measured in the same way we determined these values for the Subaru data (\S\S \ref{sec:noise} \ref{sec:comb}) and are listed in Columns 2 and 3 of Table \ref{tab:broad}.
We adopt the calibration and zero points published in \citealt{tayloretal2009}, shown here in Column 4. 

\begin{table}
 \caption{Other Optical and Infrared Data}
 \label{tab:broad}
\begin{center}
 \begin{tabular}{@{}ccccc}
 \hline
 \hline
{Band} & {FWHM [\arcsec]} &{5$\sigma$ depth [AB]\tablenotemark{1} } & {Zero Point [AB]} & {Survey} \\
{(1)} & {(2)} & {(3)} & {(4)} & {(5)}\\
 \hline
   $BVR$      &   0.83 & 26.82 & 23.58  &          MUSYC\tablenotemark{2}   \\
 \hline
    U38       &   0.98 & 25.33 & 21.96  &          GaBoDS\tablenotemark{3}   \\
    U         &   1.05 & 25.86 & 22.74  &          GaBoDS\tablenotemark{3}   \\
    B         &   1.01 & 26.45 & 24.38  &          GaBoDS\tablenotemark{3}  \\
    V         &   0.94 & 26.27 & 24.10  &          GaBoDS\tablenotemark{3} \\
    R         &   0.83 & 26.37 & 24.66  &          GaBoDS\tablenotemark{3}   \\
    I         &   0.96 & 24.30 & 23.66  &           GaBoDS\tablenotemark{3}  \\
    z         &   1.07 & 23.69 & 24.47  &          MUSYC\tablenotemark{1,4} \\
    J         &   1.48 & 22.44 & 23.53  &          MUSYC\tablenotemark{1,4}  \\
    H         &   1.49 & 22.46 & 24.15  &          ESO\tablenotemark{5}   \\
    K         &   0.94 & 21.98 & 24.40  &          MUSYC\tablenotemark{1,4}  \\
   3.6 $\mu$m &   2.08 & 23.89 & 22.42  &          SIMPLE\tablenotemark{6}\\
   4.5 $\mu$m &   2.01 & 23.75 & 22.19  &          SIMPLE\tablenotemark{6} \\
   5.8 $\mu$m &   2.21 & 22.42 & 20.60  &          SIMPLE\tablenotemark{6}\\
   8.0 $\mu$m &   2.28 & 22.50 & 21.78  &          SIMPLE\tablenotemark{6} \\
 \hline
\vspace{-0.6cm}
\footnotetext[1]{Total Magnitude}
\footnotetext[2]{\citet{gawiseretal2006lya}}
\footnotetext[3]{\citet{Hildebrandtetal2006}}  
\footnotetext[4]{\citet{tayloretal2009}}
\footnotetext[5]{\citet{moyetal2003}}
\footnotetext[6]{Damen et al. 2010, submitted}
\end{tabular}
\end{center}
\end{table}

\subsection{SIMPLE Survey}
\label{ssec:simple}
Considerable {\it Spitzer} time has been invested in the ECDF-S with both IRAC (Infrared Array Camera; \citealt{fazioetal2004}) and MIPS (the Multi-Band Imaging Photometer for {\it Spitzer}; \citealt{riekeetal2004}). 
The Spitzer IRAC/MUSYC Public Legacy in the ECDF-S (SIMPLE, Damen et al. 2010, submitted ) project obtained very deep IRAC imaging across the full ECDF-S.
The IRAC data cover $3-8$ $\mu$m, to $\sim$24th magnitude AB at $3.6 \mu$m.
The astrometry was calibrated using the MUSYC $BVR$ detection image, the same image we used in the Subaru medium-band reductions (\S \ref{sec:comb});
the resulting positional accuracy for individual sources is $\lesssim 0.3$ arcseconds ($1 \sigma$; Damen et al. 2010, submitted).
We include the IRAC data from the SIMPLE Survey in Table \ref{tab:broad}.  The values for FWHM (Column 2) and 5 $\sigma$ depth (Column 4) are measured in the same manner as for the ground-based imaging (\S\S \ref{sec:noise}, \ref{sec:comb}).  The zero point values (Column 4) come from (Damen et al. 2010, submitted).

\subsection{Astrometric Calibration}
In order to perform multi-band photometry, each image is transformed to the image plane and pixel scale (0.\arcsec267) of the stacked $BVR$ image used for detection \citep{gawiseretal2006}.  The RMS astrometric errors, compared to  an astrometric catalog of the ECDF-S (Terry Girard, private communication), are estimated to be less than 0.2\arcsec~ across the entire field \citep{gawiseretal2006}.  Using \texttt{IRAF} tasks \texttt{geomap} and \texttt{geotran} all images were resampled to the coordinate system of the $BVR$ image (both in {\it x-y} pixel coordinates and in right ascension and declination) with a North-up tangent plane projection.

\section{Photometry}
\label{sec:phot}
In order to create accurate multiwavelength SEDs, a constant fraction of light needs to be collected for each object across every band.  
If the varying point spread functions (PSFs) of each band are not taken into account, similar apertures will collect different fractions of an object's total light.
The data in our survey come from a variety of telescopes, with large variations in seeing (Tables \ref{tab:subaru} and \ref{tab:broad}, Column 2).
Experiments with the broad band images showed that failure to correct for this effect would bias the U-V colors of sources by up to $\sim$ 20\%.
Therefore, close attention was paid to obtaining accurate colors.
In this section we describe the methodology used to measure accurate colors (\S \ref{sec:psfmatch}), the selection of the apertures to be used for the photometry (\S \ref{sec:apsel}), the detection of objects on the images (\S \ref{sec:sex}) and the completeness of our detection as a function of object magnitude (\S \ref{sec:complete}).

\subsection{PSF Matching}
\label{sec:psfmatch}
We have a total of 10 ground-based broad-band images ($U$, $U38$, $B$, $V$, $R$, $I$, $z$, $J$, $H$, $K$), 4 IRAC images ($3.6$ $\mu$m, $4.5$ $\mu$m, $5.8$ $\mu$m, $8.0$ $\mu$m), and 18 medium-band images ($IA427$, $IA445$, $IA464$, $IA484$, $IA505$, $IA527$, $IA550$, $IA574$, $IA598$, $IA624$, $IA651$, $IA679$, $IA709$, $IA738$, $IA767$, $IA797$, $IA856$).  
The seeing in these images ranges from $\sim0.5$\arcsec~ to over 2\arcsec~(Tables \ref{tab:subaru} and \ref{tab:broad}).  
In this section, we describe our technique for smoothing multiple images to the same PSF to correctly measure color fluxes.

One technique commonly used to achieve uniform photometry across images in multiple filters relies on smoothing all images to the PSF size of the image with the worst seeing \citep[e.g.,][]{labbeetal2003, capaketal2007,tayloretal2009}.
Then large apertures are used for photometry, collecting flux within a uniform physical area across all images for each galaxy. 
However, degrading all images to the largest PSF size decreases the S/N significantly in each filter where the image is smoothed.
Given the large range in PSF size across our data, if we degraded all images to the worst seeing we would sacrifice the excellent image quality found in the majority of our observations.

Therefore, we apply a two-fold approach.
First, for the 12 images with narrow PSFs (Tables \ref{tab:subaru}, \ref{tab:broad}), we smooth them to the PSF of the $BVR$ $\sim 0\farcs8 $ image.
For these images, all fluxes are measured using a single aperture, which in this case provides accurate PSF-matched photometry.
Then, for the images with PSFs larger than that of the $BVR$ image (Tables \ref{tab:subaru}, \ref{tab:broad}), we degrade a copy of the $BVR$ image to match each of the larger PSF sizes.
We can then measure a color, e.g., $f_{(K)}-f_{(BVR,smoothed)}$, and scale this to the aperture flux measured in the $BVR$ 0\farcs79 image, $f_{(BVR)}$.
In other words, the PSF-matched K-band flux is measured by:
\begin{equation}
\label{colf}
f_{(K,psfmatched)}=f_{(K)} \times \frac{f_{(BVR)}}{f_{(BVR,smoothed)}}
\end{equation}
Here $f_{(BVR)}$ is the flux measured in the $BVR$ image with its native 0\farcs8 seeing, and $f_{(BVR,smoothed)}$ is the flux measured in the copy of the $BVR$ image, smoothed to have the same seeing as the $K$-band image.
Therefore, $f_{(K,psfmatched)}$ provides accurate colors when compared with the fluxes measured from the 12 images with narrow PSFs.
We note that the aperture radius in which the flux is measured is selected to be the FWHM (see \S\ref{sec:apsel}), so to measure the flux in the $K$-band image ($f_{(K)}$) and the smoothed $BVR$ image ($f_{(BVR,smoothed)}$), we use an aperture radius equal to the FWHM of the $K$-band image.
We note that in Equation 3, K represents any of the filters whose final images have PSF sizes larger than the $BVR$ image ($U$, $U38$, $B$, $V$, $I$, $z$, $J$, $H$, $K$, $IA427$,$IA445$, $IA464$, $IA505$, $IA550$, $IA574$, $IA709$, and  $3.6$, $4.5$, $5.8$, $8.0$ $\mu$m).
The final catalog contains these PSF-matched fluxes for each of the 32 bands.

\begin{figure}
\includegraphics[angle=0, width=0.47\textwidth]{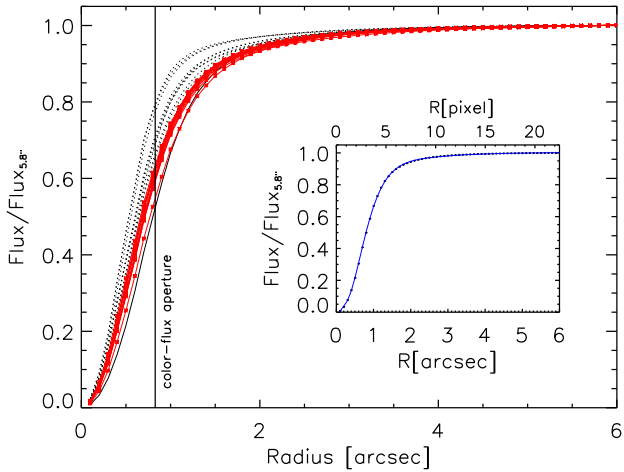}
\caption{Stellar curve of growth for the 12 bands with PSFs smaller than the $BVR$ PSF (black dashed lines), measured using 20 isolated stars.  Red solid lines show the smoothed PSF after convolution with the appropriate kernel to match the $BVR$ PSF, also indicated is the aperture used for the PSF-matched flux.  {\it Insert:} $BVR$ stellar curve of growth and fifth-order polynomial fit.  We use this curve to measure the fraction of light missed in an aperture of a given size, and thus the  correction to total flux (\S\ref{sec:sex}).}
\label{growth}
\end{figure}

Here we describe our technique for smoothing two images to an identical PSF.  
We first attempted to smooth the images with a Gaussian convolution kernel, but found that the residual PSF variations were large due to the non-Gaussian PSF shape for stars in the images.
Instead, we decided to build a separate PSF for each filter directly from the images themselves.
To do this, we selected $\sim 20$ isolated stars in each band that were well exposed but not saturated.  
The selected stars were registered, normalized to their peak flux and then averaged to create a single model PSF for each band.
We used the Lucy-Richardson algorithm (\texttt{IRAF}'s \texttt{lucy}) to construct a kernel to convolve with each image.
In Figure \ref{growth}, we show the stellar curve of growth for the 12 images with PSFs smaller than the $BVR$ image (dotted lines).
The corresponding new curves of growth for the 12 smoothed images (red lines), where we have convolved the original image with our Lucy-Richardson kernel to match the $BVR$ image PSF, show a consistent stellar profile.
The ratio of the flux enclosed within our photometry aperture to total flux (measured within 6 FWHM) is stable to nearly $\sim$ 1\%.
This accuracy is also achieved in the smoothed copy of the $BVR$ image for each filter with poor seeing.  

\subsection{Aperture Selection}
\label{sec:apsel}
In order to optimize the S/N for the photometry, the flux in each object is measured in the central high surface brightness regions of the objects and then later corrected to a total flux measurement.
For a Gaussian PSF and uncorrelated noise, the aperture diameter that maximizes the S/N is 1.35 times the seeing FWHM \citep{gawiseretal2006}.
We use an aperture radius equal to the seeing FWHM, which encloses $\sim$50\% of the flux  (Figure \ref{growth}) and provides nearly optimal S/N for point source photometry \citep{gawiseretal2006}.

We note that an aperture defined to ensure high S/N can result in fluxes that measure only the central region of galaxies, particularly for large galaxies at low redshifts.  
For example, at z=0.3, a 0$\farcs$8 aperture is $\sim$ 4 Kpc.  
Therefore, for a Milky Way size galaxy at redshift 0.3, our aperture would measure only the central bulge stars.  
Hence, any color gradients in the outer regions of the galaxy would not influence these aperture fluxes.
Therefore it is important to note that in our catalog, the fluxes describe only the central regions of the galaxies at lower redshift.
We also emphasize the importance of measuring the same central fraction of each galaxy in the apertures (\S \ref{sec:psfmatch}).

When measuring the noise in the image (\S~\ref{sec:noise}), we used an aperture size equal to the seeing FWHM and converted these to total magnitudes assuming a stellar profile (Figure \ref{growth}).
Therefore our 5$\sigma$ depths are computed for point sources, a magnitude offset of -0.6 from the aperture depth derived from the images.

\subsection{Source Detection}
\label{sec:sex}
For object detection we run SExtractor version 2.4.4 \citep{bertinarnouts1996} in dual-image mode using the original BVR image for detection.
In the 20 filters where a copy of the BVR image was smoothed to the image's larger PSF, SExtractor was run twice: first on the filter's image and then again on the smoothed BVR image (\S \ref{sec:psfmatch}).
SExtractor directly gives us aperture fluxes (with a radius of 2 x FWHM) for the images of the 12 filters with small PSFs.
For the 20 filters with images with larger PSFs, we compute aperture fluxes from Equation \ref{colf}. 

To correct our aperture fluxes to total fluxes for each source, we compute a total-flux correction factor for each object.
SExtractor's AUTO flux uses a flexible Kron-like \citep{kron1980} elliptical aperture to compute the flux.
The flux in this aperture accounts for the size of the source and the source shape, and measures approximately 94\% of the total flux \citep{bertinarnouts1996}.
We use SExtractor to obtain AUTO fluxes for each object from the BVR image. 
Then we correct this AUTO flux to a total flux using the stellar curve of growth from the BVR image (Figure \ref{growth} insert, points).
We fit this growth curve with a 5th order polynomial (Figure \ref{growth} insert, solid blue line) and compute the fraction of light enclosed for each object at the Kron radius ($lightfrac$).  
The total flux correction factor to be applied to the AUTO flux is then the inverse of the fraction of light enclosed at the AUTO flux radius for each source ($totcor=1/lightfrac$).
To correct our aperture fluxes in any given filter to total flux in that filter we scale the aperture flux using the ratio of the BVR AUTO flux to the BVR aperture flux for that source and then multiply it by the flux correction factor ($totcor$):
\begin{equation}
\label{totcor}
f_{K,tot}=f_{K,aperture}\times \frac{f_{BVR,AUTO}}{f_{BVR,aperture}} \times {totcor},
\end{equation}
where $K$ can be replaced here by any filter in our catalog.
We note that for extended objects this is a minimum correction to a total magnitude \citep{tayloretal2009}.

\subsection{Completeness}
\label{sec:complete}
Although we measure the nominal depth of our images using the background noise (\S \ref{sec:noise}) in empty apertures across the field, the actual completeness of our catalog is a strong function of source magnitude.
For each filter in the catalog, the 5$\sigma$ depth is listed in Tables \ref{tab:subaru} and \ref{tab:broad}; for the BVR detection image, this depth is 26.8 in AB magnitudes.

\begin{figure}
\includegraphics[angle=0, width=0.47\textwidth]{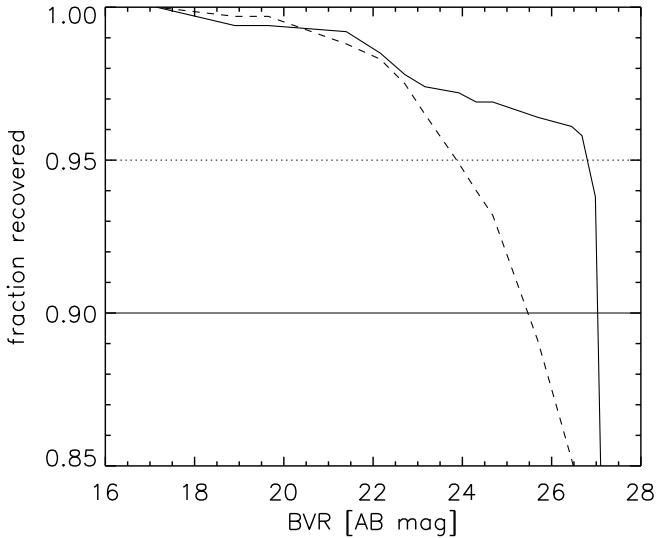}
\caption{The fraction of simulated stars detected by Sextractor.  For stars placed to avoid bright stars and the locations of other objects on the image (solid line), the 90\%/95\% completeness level is nearly 27 AB magnitudes.  Stars placed randomly (dashed line), yield lower completeness because they suffer from source confusion.  The 90\% (95\%) completeness for these simulated stars is $\sim$25.5 (24) AB magnitudes.}
\label{fig:complete}
\end{figure}

To empirically measure our overall effectiveness in detecting objects as a function of magnitude, we construct a stellar PSF, normalize it to various magnitudes, place it into the BVR detection image and determine if SExtractor recovers the inserted star.
The stellar image we use is that constructed from 20 well exposed stars in Section \ref{sec:psfmatch} to measure the BVR image PSF.
A thousand random locations are selected, at locations at least 10 pixels ($\sim$3 FWHM) away from detected objects and avoiding regions of the field washed out by bright stars.  
Near larger objects (eg. galaxies), distances to these random sources were increased to be greater than twice the Kron radius.
We create multiple images, each with 1000 stars of varying magnitude and then run SExtractor on each of these new images.
Figure \ref{fig:complete} shows the percentage of these stars recovered as a function of simulated magnitude (solid line).
The 95\% completeness level is nearly 27 mag, and it falls quickly below that magnitude.
This agrees well with the number determined from placing empty apertures around the field.  
However, in both cases we ignore the extent of detected objects on the field.
Because real galaxies do overlap each other along the line of sight, we repeat this exercise, no longer restricting the random positions to avoid detected objects of this field (although we still avoid the very bright stars which wash out large areas on the field).
The recovery rate of these sources falls much faster at faint magnitudes, due to missing the inserted stars near brighter stars and galaxies (Figure \ref{fig:complete}, dashed line).
Still, at a magnitude of $\sim$25.5 we are 90\% complete in our detections.

\begin{figure}
\includegraphics[angle=0, width=0.47\textwidth]{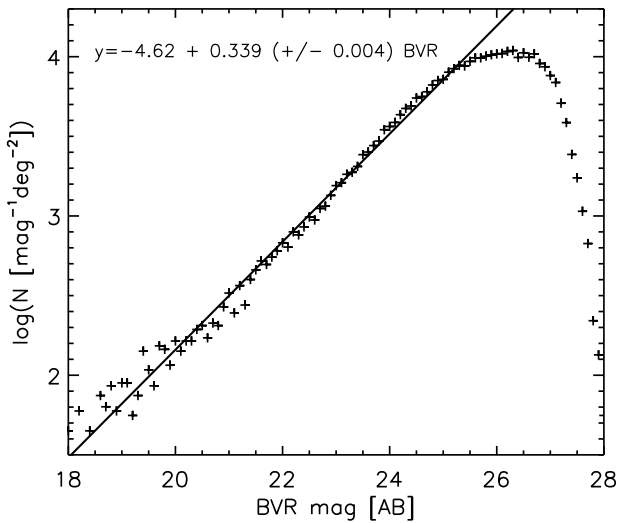}
\caption{Number counts in the BVR-band.  The number of sources increases steadily to BVR$\sim$ 25.5 mag, our 90\% completeness level.  Our differential number count measurement 0.339 $\pm$ 0.004  consistent with the value measured by \citep{gawiseretal2006} 0.34  $\pm$ 0.01. }
\label{rcts}
\end{figure}

 Figure \ref{rcts} shows our $BVR$ band number counts for the BVR-dected catalog. The number counts increase steadily until BVR$\sim$25.5 mag [AB], where we are still 90\% complete.  Differential number counts can measure the geometry of space and the evolution of structure in the universe.  Our fit to the differential number counts per magnitude per square degree is 0.339 $\pm$ 0.004, consistent with the 0.34 $\pm$ 0.01 measured by \citet{gawiseretal2006}.  

\subsection{Galactic Extinction}
\label{sec:gext}
The ECDF-S is at high Galactic latitude and therefore has a very low Galactic extinction.
For the location of the ECDF-S, RA=3$^h$ 32\arcmin~ \& DEC=-27\arcdeg 48\arcmin, we calculate a value of $E(B-V)=0.0088$ from the 100 $\mu$m maps of  \citet{schlegeletal1998}\footnote{http://www.astro.princeton.edu/~schlegel/dust/data/data.html}.
We calculate the expected Galactic extinction in each band, assuming R=3.1 and using the Galactic Extinction Curve of \citet{cardelli1989} with updates in the optical region from \citet{odonnell1994}.
These values are listed in Table \ref{tab:photoff}, Column 1.
The Galactic extinction corrections are not included in the catalog, but are applied before photometric redshifts are computed (\S\ref{sec:redshifts}).

\subsection{Catalog}
\label{sec:cat}
We present our catalog format in Table \ref{tab:cat}, available online in full format\footnote{\texttt{http://physics.rutgers.edu/\~{}gawiser/MUSYC}}.  
The photometry is measured in units of flux ($\mu$Jy) and is not corrected for Galactic extinction.  
For the $BVR$ combined image, the \texttt{AUTO} flux as well as the \texttt{APERTURE} flux is included.
The fluxes are aperture fluxes and can be corrected to total fluxes using the total flux correction and the AUTO flux (Equation \ref{totcor}).
The geometrical parameters are output from Sextractor including the Kron Radius, \texttt{A\_IMAGE}, \texttt{B\_IMAGE},\texttt{THETA\_IMAGE} and \texttt{CLASS\_STAR}.
The table is presented both as a text file and as a FITS file.  In the FITS version of the catalog, SExtractor detection flags \citep{bertinarnouts1996} for each filter are included. 

 \begin{table*}
 \caption{Summary of Photometric Catalog Contents}
 \label{tab:cat}
\begin{center}
 \begin{tabular}{@{}lll}
 \hline
 \hline
{Column No.} & {Column Title} &{Description} \\
 \hline
1 	& num 			& Squential Object Identifier, beginning from 0 \\
2,3 	& ra, dec	 		& right ascension \& declination (J2000; decimal degrees)\\
4 	& CLASS\_STAR 	& SExtractor parameter measuring stellarity of object \\
5 	& Kron radius 		& Sextractor parameter measuring source size in a flexible aperture (pixels)\\
6,7  	& A\_IMAGE,B\_IMAGE& SExtractor parameter measuring major and minor axis of image profile (pixels)\\
8	& THETA\_IMAGE 	& SExtractor parameter measuring  position angle measured counterclockwise from North (degrees) \\
9 	& totcor 			& aperture correction to convert AUTO flux to total ($\mu$Jy)\\
10,11	& f\_auto\_BVR,e\_auto\_BVR & SExtractor AUTO flux and error ($\mu$Jy) \\
12-76 & f\_X,e\_X & aperture flux and error in each filter ($\mu$Jy), including that measured for the BVR image\\ 
77 &flag\_X			& BVR SExtractor detection flag *\\
\hline
\end{tabular}
{*note: in the FITS version of the table, SExtractor detection flags are included for all filters.}
\end{center}
\end{table*}

\section{Redshifts}
\label{sec:redshifts}
The Subaru imaging allows us to obtain accurate ($\Delta{\rm z/(1+z)} \sim 0.01$) photometric redshifts for the sources in our catalog.  In this section we describe our method for computing photometric redshifts and evaluate their accuracy as determined by a subset of sources with spectroscopic redshifts.

\subsection{EAzY}
\label{sec:eazy}
In order to obtain highly accurate photometric redshifts, we used EAzY, a program optimized to provide high quality redshifts over 0 $\le$ z $\le$ 4, where complete spectroscopic calibration samples are not available \citep{brammeretal2008}.  
EAzY is a full-featured redshift fitting code, allowing for the use of priors in  computing photometric redshifts (e.g., BPZ; \citealt{benitez1999}).
It includes a user-friedly interface based on HYPERZ \citep{bolzonellaetal2000} and a carefully selected template set, designed to optimize photometric redshifts for optical-NIR surveys \citep{brammeretal2008}.  
The template set and the magnitude priors are based on semi-analytical models that are complete to very faint magnitudes, rather than highly biased spectroscopic samples, and so are particularly useful for samples of objects such as dust obscured galaxies that are faint in the optical \citep[e.g.,][]{treisteretal2009dog}.
When running EAzY, we allow for a linear combination of all templates and include a broad-line AGN template only when fitting sources detected in X-rays \citep{cardamoneetal2008}.  
The AGN template is based on the SDSS QSO template from \citet{vandenberketal2001}, which we've extended towards the red using the mean QSO SED from \citet{richardsetal2006}.
Following \citet{Ilbertetal2009}, EAzY now includes strong narrow emission lines in its galaxy templates by estimating an [\ion{O}{2}] emission line flux from the UV luminosity of the scaled template and adopting fixed line ratios for [\ion{O}{3}/\ion{O}{2}], [H$\beta$/\ion{O}{2}], and [H$\alpha$/\ion{O}{2}] as defined by \citet{kennicutt1998}.
Further, EAzY introduces a template error function to account for wavelength-dependent template mismatch.

EAzY provides multiple estimates of the photometric redshift, including z\_peak, which we adopt in this work as the photometric redshift estimate.
Because z\_peak is marginalized over the full probability distribution, it can differ from the best estimate from a straight $\chi^2$ minimization when there are two widely-separated peaks in the redshift probability function by selecting the peak with the largest integrated probability.
Additionally, it includes an estimate of the quality of each photometric redshift ($Q_z$) which combines the $\chi^2$ of the template fit, the width of the 68\% confidence interval and the BPZ odds parameter \citep{benitez1999} in such a way that $Q_z$ increases as any of those parameters deteriorates \citep{brammeretal2008}.
We recommend adopting a cut in $Q_z$, when using the photometric redshifts for science (see \citealt{brammeretal2008} and \S\ref{sec:speczphot}), and in this work adopt the conservative requirement of $Q_z$ $\le$ 1 for all photometric redshifts used.

\begin{figure*}
\includegraphics[angle=0, width=0.99\textwidth]{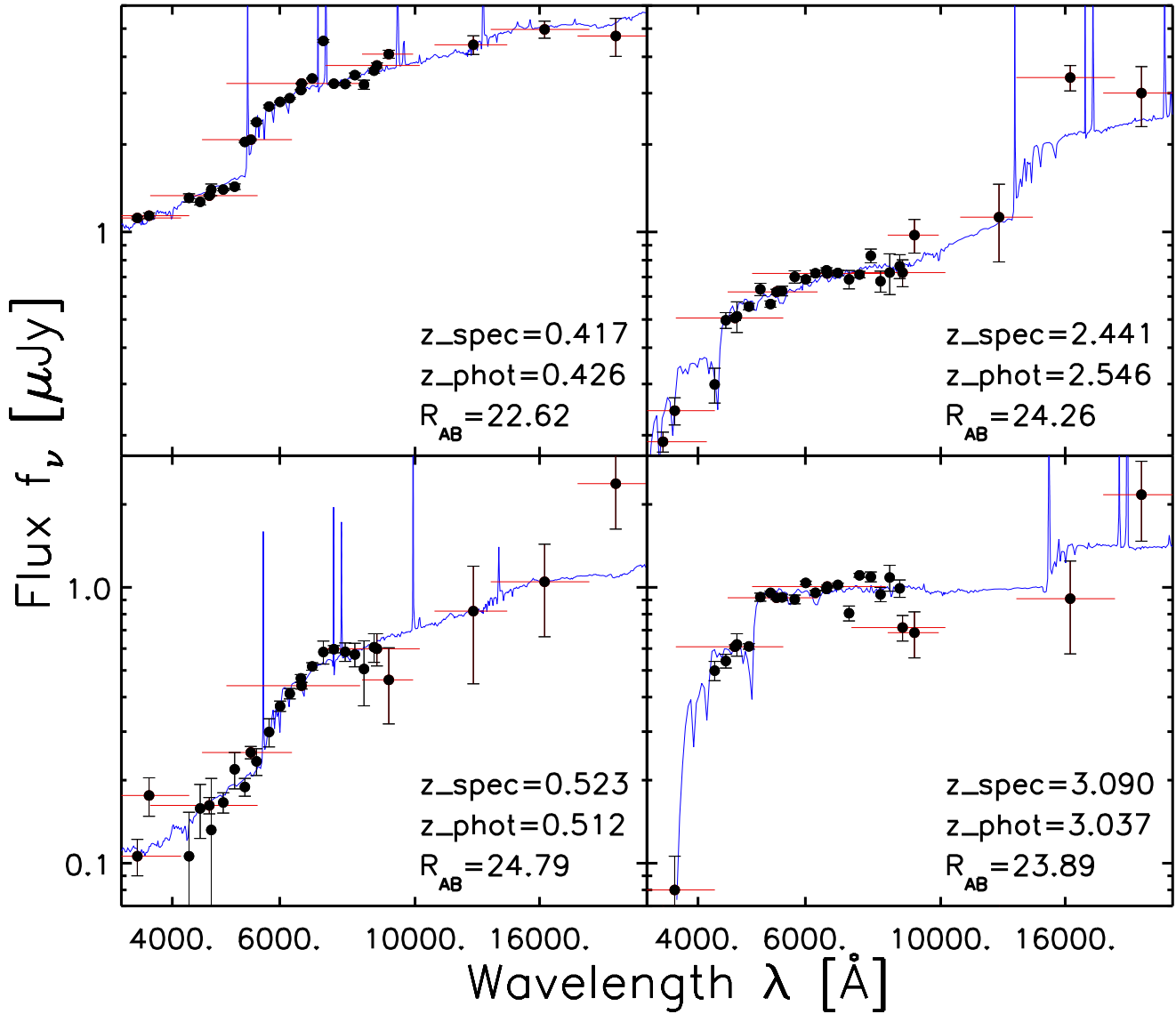}
\caption{Example SEDs with EAZY spectral fits overlayed.  In black are the data points with 1$\sigma$ errors.  The blue line shows the linear combination of template SEDs that best fit the data. In the left hand column, we illustrate two low redshift examples where the Balmer break is fit by the medium-band filters.  In the right hand column, we illustrate two high redshift examples showing a break due to Lyman-alpha absorption.   We highlight fits to faint sources, R $\sim$ 24 in 3 of the 4 sources shown.  Emission lines do influence the photometry. }
\label{fig:seds}
\end{figure*}

When using EAZY to compute our photometric redshifts, we include the 18 medium bands and the optical and near-infrared ground based coverage in addition to the IRAC data.
Example SEDs are shown in Figure \ref{fig:seds}.
The FWHM of the broad band filters is indicated by a red line.
We compute photometric redshifts for over 99\% of sources in our BVR-detected catalog using the default EAZY output, and adopting the default template error function and R-band photometric prior. 
For the less than 1\% of BVR-detected sources that lie on the imaging area in fewer than 5 filters, we do not compute photometric redshifts.

\subsection{Spectroscopic Redshifts}
\label{sec:specz}
We collect all available spectroscopic redshifts for sources in our catalog from the literature in order to quantify our photometric redshift accuracy.  
The quality of these redshifts varies widely, from sources with multiple spectral line measurements to sources showing only hints of a single spectral line.  
Matching these spectroscopic redshifts to our catalog, there are almost $\sim$4000 unique objects with redshifts, $\sim$1000 of  which have multiple published spectroscopic redshifts. 
For sources with multiple redshift determinations we select those with a higher quality flag, favoring redshift measurements with published quality flags over those without.  
In Table \ref{tab:specz}, we list all references from which unique spectroscopic redshifts were obtained, including the number of spectroscopic redshifts used from the data-set, the source paper, the quality flags we used in determining the accuracy of our photometric redshifts and the median R-band magnitude (\S~\ref{sec:speczphot}).
When evaluating the accuracy of the photometric redshifts, we restrict the quality of the spectroscopic redshifts used in the comparison. 
If we include spectroscopic redshifts with lower quality flags, it lowers the determined accuracy by factors of 2 or 3, independent of redshift or magnitude, and it increases the fraction of outliers. 
Each of the redshift catalogs adopted a different selection technique for the sources for which they obtained spectra: 
\citet{balestraetal2010} selected galaxies at $1.8 < z < 3.5$ for the VIMOS Low Resolution Blue grism and galaxies at $z <1$ in addition to Lyman Break Galaxies at $z > 3.5$ in the Medium Resolution (MR) orange grism, 
\citet{vanzellaetal2008} selected galaxies using color criteria and photometric redshifts at $0.5 < z < 2$ and $3.5 < z < 6.3$ for VLT/FORS2 observations,
\citet{lefevreetal2004} selected galaxies using a magnitude selection of $\rm I_{AB} < 24$,
\citet{cimattietal2002} selected galaxies using a magnitude selection of $\rm K_s < 20$,
Lira et al., in prep, selected galaxies using a variety of criteria including sources with X-ray-counterparts, Lyman Alpha Emitters and Lyman Break galaxies at $2.7 < z < 3.6$,
\citet{szokolyetal2004} selected galaxies with X-ray-counterparts,
\citet{krieketal2008} selected K-bright galaxies at $z\sim2.3$ using the BzK and DRG color selection \citep{daddietal2004, franxetal2003},
\citet{treisteretal2009} selected galaxies with X-ray-counterparts,
\citet{strolgeretal2004} selected high redshift galaxies in a Supernova Search,
 \citet{cristianietal2000} selected galaxies using the VLT UV-Visual Echelle Spectrograph to detect Ly$\alpha$,
 \citet{vanderweletal2004} selected galaxies at $z\sim1$ from the COMBO-17 catalog with a compact and regular shape,
and  \citet{croometal2001} selected compact objects with J-K colors redder than the stellar sequence. 
We note that many of these spectroscopic datasets have a larger number of total published redshifts; we include here only those which we have adopted as unique redshifts in our spectroscopic redshift catalog.  
Aditionally, other references contain spectroscopic redshifts in the ECDF-S, but they are not mentioned here unless we adopt at least one spectroscopic redshift into our unique listing.

\begin{table}
 \caption{Spectroscopy}
 \label{tab:specz}
\begin{center}
 \begin{tabular}{@{}rlrc}
 \hline
 \hline
{\# Sources} & {References} & {Quality Flags} & {median($\rm R_{AB}$)}\\
{(1)} & {(2)}  & {(3)} & {(4)} \\
 \hline  
 1239 & \citealt{balestraetal2010} & A & 23\\
  573 & \citealt{vanzellaetal2008} & A,B & 24\\
  223 & \citealt{lefevreetal2004} & 4,3 & 23\\
  224 & \citealt{cimattietal2002} & 1,0 & 23\\
  211 & Lira et al., in prep & 3,2 & 22\\
  52 & \citealt{szokolyetal2004} & 3,2,1 & 23\\
  9 & \citealt{krieketal2008} & n/a & 24\\
  7 & \citealt{treisteretal2009} & 1 & 23\\
  5 & \citealt{strolgeretal2004} & n/a & 25\\
  3 & \citealt{cristianietal2000} & n/a & 24.5 \\
  3 & \citealt{vanderweletal2004} & n/a & 26\\
  2 & \citealt{croometal2001} & n/a & 22\\
\hline
\end{tabular}
\end{center}
\end{table}

\subsection{Zero Point Adjustments}
\label{sec:photoff}
To perform SED fitting, we require highly accurate colors across the optical and near-infrared spectrum.
Photometric zero points (Tables \ref{tab:subaru} and \ref{tab:broad}, Column 4) were determined individually in each band through standard star measurements.
However, small offsets of a few percent in adjacent bands can introduce significant color-offsets.
This is a particular concern when a catalog, like ours, includes photometry from a variety of instruments taken over a period of many years.
If we assume that {\it apriori} we know the spectral shapes of the galaxies and that they can be well fit by our synthetic SED templates, than any systematic offsets of observed fluxes in a given filter are due to a  photometric zeropoint error in that filter.
Because our sources cover a wide range of redshifts, each filter samples a different region in every rest-frame galaxy SED and therefore any systematic photometric offsets in this filter are due to zero point offsets rather than a template error at a single wavelength.

To cross-calibrate our photometry, we take the best-fit EAzY SED for each galaxy and measure the offset of observed flux from that template flux in each filter.
For normally distributed errors in flux measurements, the average of these offsets should be zero.
However, we find systematic flux offsets between the observed flux and the template flux for each filter.
We wish to correct the filters zero point by the median offset in each filter.
For this calculation we restrict our sample to sources with spectroscopic redshifts and high S/N detections, in this case expected SED template fluxes of at least 5 $\times$ the detection limit for the filter.
We calculate the median flux offset for each filter, iterating by applying the calculated flux offsets and re-fitting the SEDs.
Within 3 iterations, these flux offsets vary by less than 1\%, and we find that slightly altering the list of sources used to calculate these flux offsets does not affect our solutions by more than 1-2\%.
Therefore our photometry is systematically uncertain at this level.
For each band, we apply these photometric offsets, listed in Table \ref{tab:photoff}, before computing final photometric redshifts used in this paper.
We do not calculate offsets for the IRAC bands where the templates have greater uncertainties.

\begin{table}
 \caption{Photometric Zero Point Offsets from SED Fitting}
 \label{tab:photoff}
\begin{center}
 \begin{tabular}{@{}rcc}
 \hline
 \hline
 {Band} & {Galactic Extinction} & {ZP Correction} \\
{(1)} & {(2)}   &{(4)} \\
 \hline
  U38 & 0.041  & -0.184  \\
    U & 0.043  & -0.245  \\
    B & 0.034  &  0.053  \\
    V & 0.028  &  0.075  \\
    R & 0.022  &  0.075  \\
    I & 0.014  &  0.032  \\
    z & 0.013  &  0.156  \\
    J & 0.008  & -0.023  \\
    H & 0.005  & -0.011  \\
    K & 0.003  &  0.255  \\
IA427 & 0.037  &  0.285  \\
IA445 & 0.036  &  0.124  \\
IA464 & 0.034  & -0.057  \\
IA484 & 0.032  &  0.075  \\
IA505 & 0.030  &  0.146  \\
IA527 & 0.029  &  0.010  \\
IA550 & 0.027  &  0.126  \\
IA574 & 0.026  &  0.054  \\
IA598 & 0.024  &  0.064  \\
IA624 & 0.023  &  0.116  \\
IA651 & 0.022  &  0.032  \\
IA679 & 0.021  & -0.035  \\
IA709 & 0.020  & -0.001  \\
IA738 & 0.019  &  0.054  \\
IA767 & 0.018  &  0.086  \\
IA797 & 0.017  &  0.075  \\
IA827 & 0.016  & -0.106  \\
IA856 & 0.014  &  0.106  \\
\hline
\end{tabular}
\end{center}
\end{table}

\subsection{Flux Comparison to other Catalogs}
The region around the MUSYC-Extended Chandra Deep Field-South has been observed by many surveys and therefore there are many public catalogs available in this region.
We refer to \citet{tayloretal2009} for a detailed list of these surveys and a quantitative comparison of their photometry.
Here we briefly compare our catalog to two previous versions of the MUSYC catalog, from \citet{gawiseretal2006} and \citet{tayloretal2009}, and to the updated photometry of the Combo-17 medium-band catalog, \citep{wolfetal2008}.
Each of these catalogs used slightly different techniques to derive the PSF-matched photometry, but all used Sextractor for detection and the same publicly available data.
On average our fluxes are within 1\% of those quoted by \citet{gawiseretal2006}.
For the comparison to \citet{tayloretal2009} and \citet{wolfetal2008}, we include the offsets calculated in \S \ref{sec:photoff}.
Our photometry agrees with that published by \citet{tayloretal2009} on average to $\sim$ 0.02 magnitudes, and with that of Combo-17 \citep{wolfetal2008} with an offset of -0.05 magnitudes.

\subsection{Photometric Redshifts}
\label{sec:speczphot}
\begin{figure*}
\includegraphics[angle=0, width=0.5\textwidth]{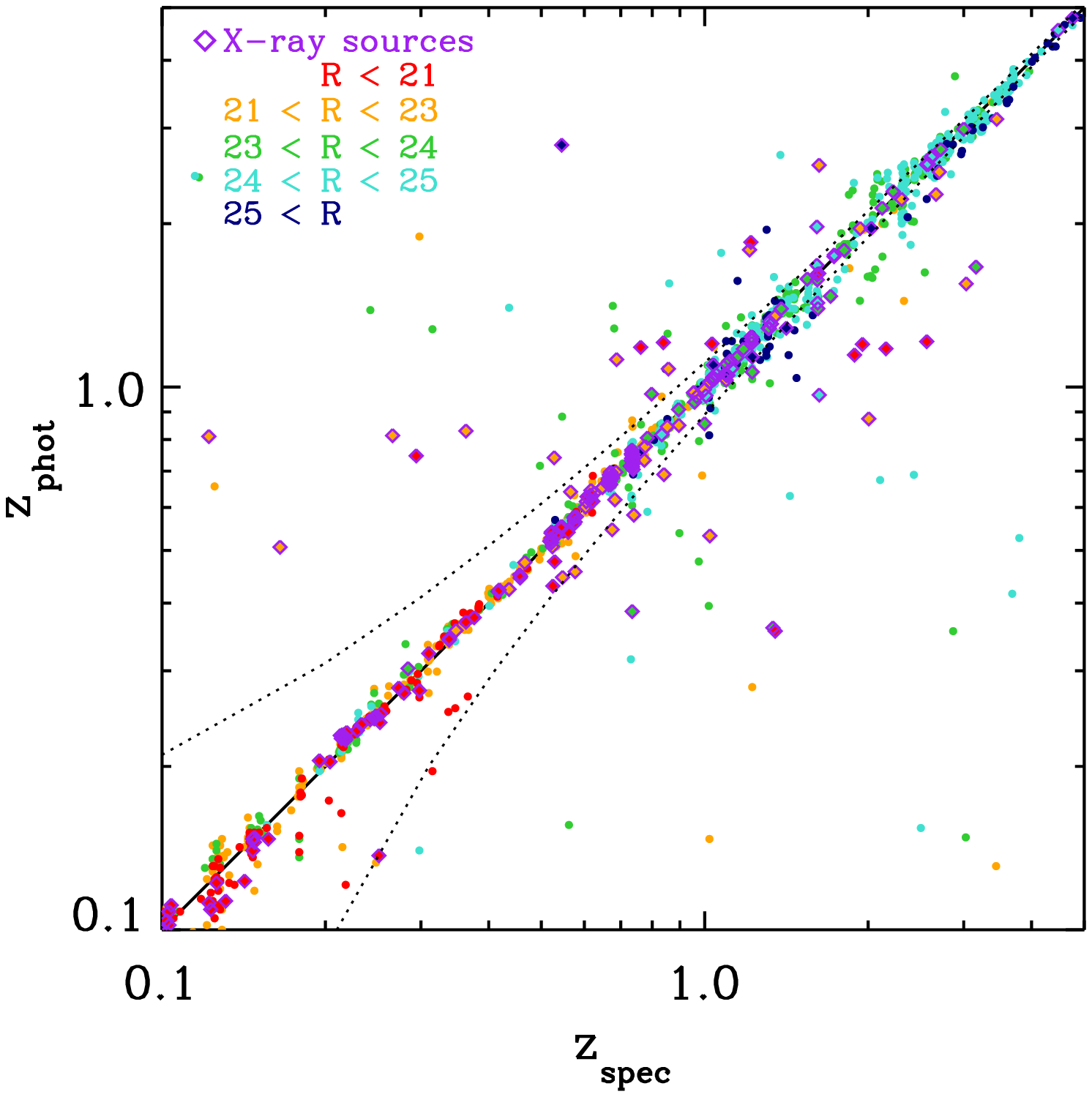}
\includegraphics[angle=0, width=0.5\textwidth]{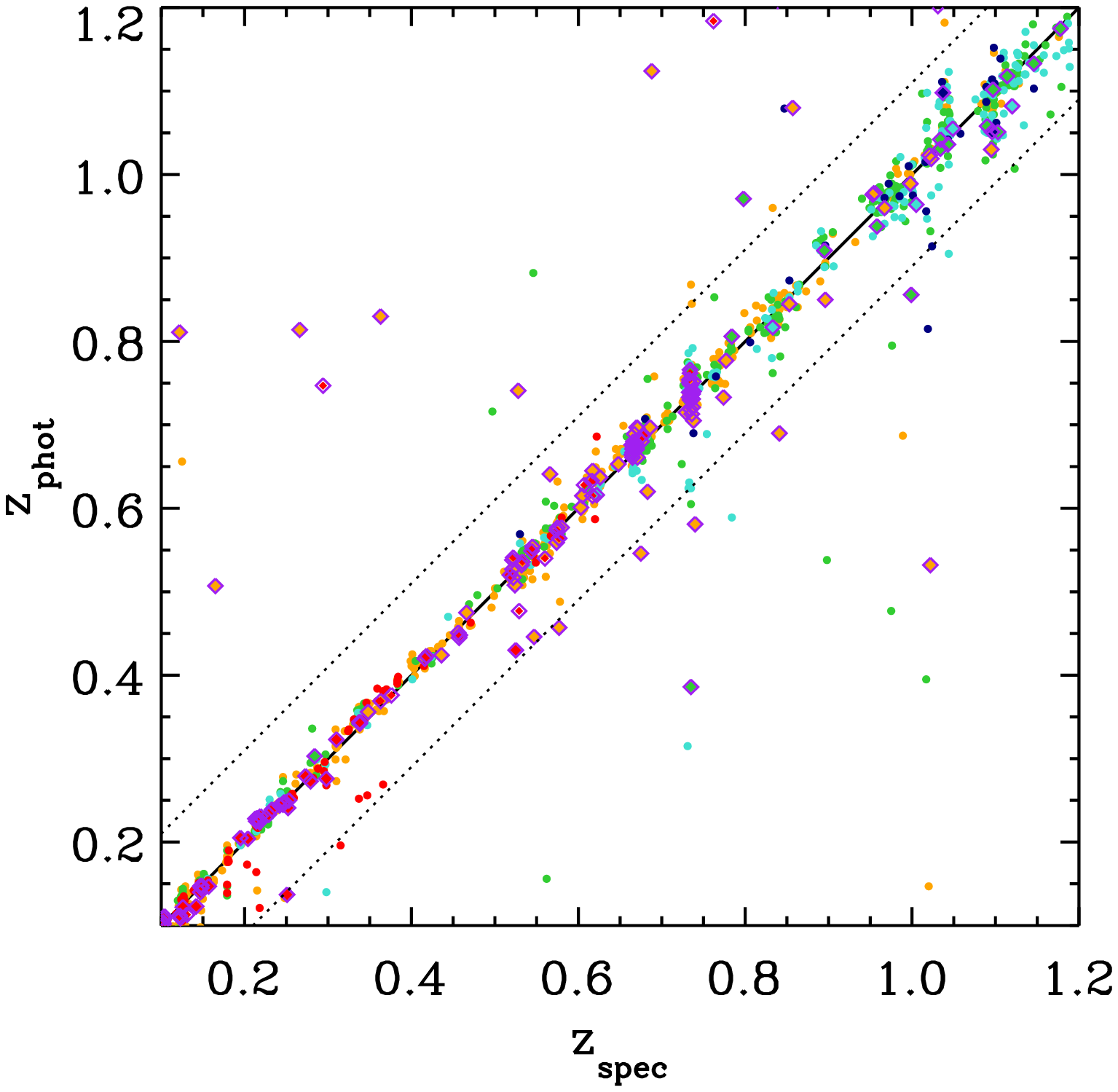}
\caption{{\it Left:} Comparison of photometric and spectroscopic redshifts for all high quality spectroscopic redshifts (Table \ref{tab:specz}) and photometric redshift fits with $Q_z \le 1$.  The galaxies are colored by R-band magnitude, and those detected in X-rays are also indicated by purple diamonds.  We find $\Delta{\rm z/(1+z)} \sim 0.008$, or $\Delta{\rm z/(1+z)} \sim 0.006$ at $0.1\le$z$\le$1.2, where the Balmer break falls into the wavelength range covered by our medium-band filters. Dotted lines are set at 10$\times$NMAD, our definition of outliers.  {\it Right} Zoom-in on low redshift region with linear scale. }
\label{eazycomp}
\end{figure*}

\begin{table*}
 \caption{Photometric Redshift Quality vs Source Magnitude}
 \label{tab:photzvmag}
\begin{center}
 \begin{tabular}{@{}lrr c r cc r}
 \hline
 \hline
  {Magnitude}  & {\# Objects} & {$Q_z\le1$} & {${\rm l68_z-u68_z}$} & {\# $z_{spec}$} & {$\Delta{\rm z/(1+z)}$} & {$\sigma_{\rm z}$} & {outliers} \\
 \hline
17 $<$ R $<$ 19   &    283   & 99\% &  0.011   &     20  &   0.004  &   0.005  &  5\% \\
19 $<$ R $<$ 20   &    385   & 98\% &  0.012   &     32  &   0.005  &   0.006  &  3\%\\
20 $<$ R $<$ 21   &    733   & 98\% &  0.014   &     81  &   0.006  &   0.008  &  1\% \\ 
21 $<$ R $<$ 22   &   1676   & 97\% &  0.015   &    167  &   0.006  &   0.007  &  2\% \\
22 $<$ R $<$ 23   &   3537   & 93\% &  0.017   &    374  &   0.006  &   0.008  &  2\% \\
23 $<$ R $<$ 24   &   8087   & 86\% &  0.022   &    524  &   0.007  &   0.011  &  6\% \\  
24 $<$ R $<$ 25   &  17048   & 68\% &  0.038   &    415  &   0.014  &   0.020  &  5\% \\
25 $<$ R $<$ 26   &  25628   & 31\% &  0.063   &     77  &   0.019  &   0.032  &  6\% \\    
26 $<$ R        &  26127   &  6\% &  0.121   &      7  &   0.021  &   0.026  & 14\% \\    
\hline
 X-ray Sources         &    825   & 69\% &  0.022   &    236  &   0.008  &   0.012  & 12\% \\    
 
\hline
\end{tabular}
\end{center}
\end{table*}
\begin{table*}
 \caption{Photometric Redshift Accuracy vs Redshift}
 \label{tab:photzvsz}
\begin{center}
 \begin{tabular}{@{}l r c c r cc r}
 \hline
 \hline
  {$z_{phot}$}  & {\# $Q_z\le1$} &{${\rm l68_z-u68_z}$} &   {$Q_z\le1$ (\% of $z_{spec}$)} & {\# $z_{spec}$} & {$\Delta{\rm z/(1+z)}$} & {$\sigma_{\rm z}$} & {outliers} \\
 \hline
~ 0 $<$ z           &  31381 &  0.034   &   92\% &1697  &   0.008  &   0.011 &   4\%   \\
~ 0 $<$ z $<$ 1.2   &  22318 &  0.024 &98\%  &   1242  &   0.006  &   0.008 &   3\%   \\
1.2 $<$ z $<$ 3.7   &   8547 &  0.087   &80\% &    433  &   0.019  &   0.027 &   8\%   \\
3.7 $<$ z           &    516 &  0.065        &92\% & 21  &   0.012  &   0.016 &   5\%   \\
 \hline
\end{tabular}
\end{center}
\end{table*}
Comparing non-X-ray sources with high quality spectroscopic redshifts, we find a median accuracy $\Delta{\rm z/(1+z)} \sim 0.008$\footnote{\rm $\Delta z$=$|z_{spec}-z_{phot}|$} out to z$\sim$5 (Figure \ref{eazycomp}, {\it left}). 
Limiting ourselves to the subsample of sources with photometric redshifts $0.1\le z\le1.2$, where the Balmer break falls within our medium bands, we find $\Delta{\rm z/(1+z)} \sim 0.006$ (Figure \ref{eazycomp}, {\it right}).
We report the NMAD \footnote{We calculate the normalized median absolute deviation (NMAD) as $$\sigma_{\rm NMAD} =  1.48\times {\rm median \left| \frac{ \Delta{\rm z-median}(\Delta z)}{1+z_{spec}} \right|}$$ as in \citep{brammeretal2008}. The normalization factor of 1.48 ensures the NMAD of a Gaussian distribution is equal to its standard deviation and the subtraction of median($\Delta z$) corrects any offset from zero of the Gaussian. The NMAD is a useful measure of dispersion because it is less sensitive to outliers that the standard deviation \citep{Ilbertetal2006,brammeretal2008}} 
for comparison with the works of others (e.g., \citealt{Ilbertetal2006,brammeretal2008,tayloretal2009}), and refer to it with the symbol $\sigma_{\rm z}\sim0.01$.
We define outliers as 10$\times\sigma_{\rm z}$, and find that $\sim$5\% of the overall sample are labeled as outliers.

In Table \ref{tab:photzvmag}, we show the photometric redshift quality as a function of galaxy magnitude; listing the magnitude bin, number of objects found within the given magnitude bin,  the percent of the galaxies in the bin with $Q_z \le 1$, the median 68\% confidence interval for the galaxies in the bin as determined by EAzY, the number of galaxies in the bin with high quality spectroscopic redshifts, the median $\Delta{\rm z/(1+z)}$, $\sigma_{\rm z}$ and the percent of galaxies in the bin with spectroscopically confirmed redshifts that are outliers.  
The number of high quality comparison spectroscopic redshifts is a strong function of magnitude, but overall we maintain median $\Delta{\rm z/(1+z)} \la 0.01$ ($\sigma_{\rm z} \la 0.02$) to an R-band magnitude of 25. 
The scatter in the photometric vs. spectroscopic redshifts almost certainly underestimates the actual uncertainty in the redshifts \citet{brammeretal2008}, and therefore we use the median 68\% confidence intervals computed by EAzY to compare the quality of the photometric redshifts for all galaxies.
This confidence interval is the difference between the 68\% confidence upper and lower bounds on the photometric redshift computed from the probability distribution $p(z)$ \citep{brammeretal2008} and is a strong function of galaxy apparent magnitude.
We find good quality photometric redshifts ($Q_z$ $\le$ 1; \citealt{brammeretal2008}) down to R-band magnitude of 25 for 70\% of the sample (90\% of the sample has $Q_z \le 3$), but miss a large fraction of the galaxy population at fainter magnitudes.
For many of the remaining sources poor fits are obtained by EAzY because the photometry is too uncertain, due to faintness of the sources, or intrinsic variability in the source over the time period of which the photometry was taken \citep{salvatoetal2009}.  
There are also cases where the intrinsic SED may not be matched by the templates and/or degeneracies in color-z space result in multiple peaks in the redshift-probability distribution.
We note that when considering X-ray counterparts (Figure \ref{eazycomp}, purple diamonds), our accuracy is maintained $\Delta{\rm z/(1+z)} \sim 0.009 $, but the outlier fraction increases to 18\%.
We include the X-ray counterparts in the last line of Table \ref{tab:photzvmag} for comparison.

Additionally, we compare the photometric redshift accuracy as a function of redshift in Table \ref{tab:photzvsz}.
Because the spectroscopic sample contains a variety of selection effects (\S \ref{sec:specz}),  Table \ref{tab:photzvsz} includes the median 68\% confidence intervals computed by EAzY for all objects with $Q_z$ $\le$ 1.
We use photometric redshift to select the sources in each redshift bin so that we are comparing comparable sources in our measurements of ${\rm l68_z-u68_z}$ and $\sigma_{\rm z}$.
Comparing to the galaxies with spectroscopically determined redshifts, the completeness of the photometric redshift determinations ($Q_z$ $\le$ 1) varies as a function of redshift.  
We fit high quality photometric redshifts to over 90\% of galaxies, but we note that the completeness falls to 80\% in the interval between $1.2 < z < 3.7$.
Overall, with medium-band photometry we have achieved highly accurate photometric redshifts for the majority of the sources in the ECDF-S.
The photometric redshifts are available online through the MUSYC webpage\footnote{\texttt{http://physics.rutgers.edu/\~{}gawiser/MUSYC}}. 

\begin{figure*}
\includegraphics[angle=0, width=0.99\textwidth]{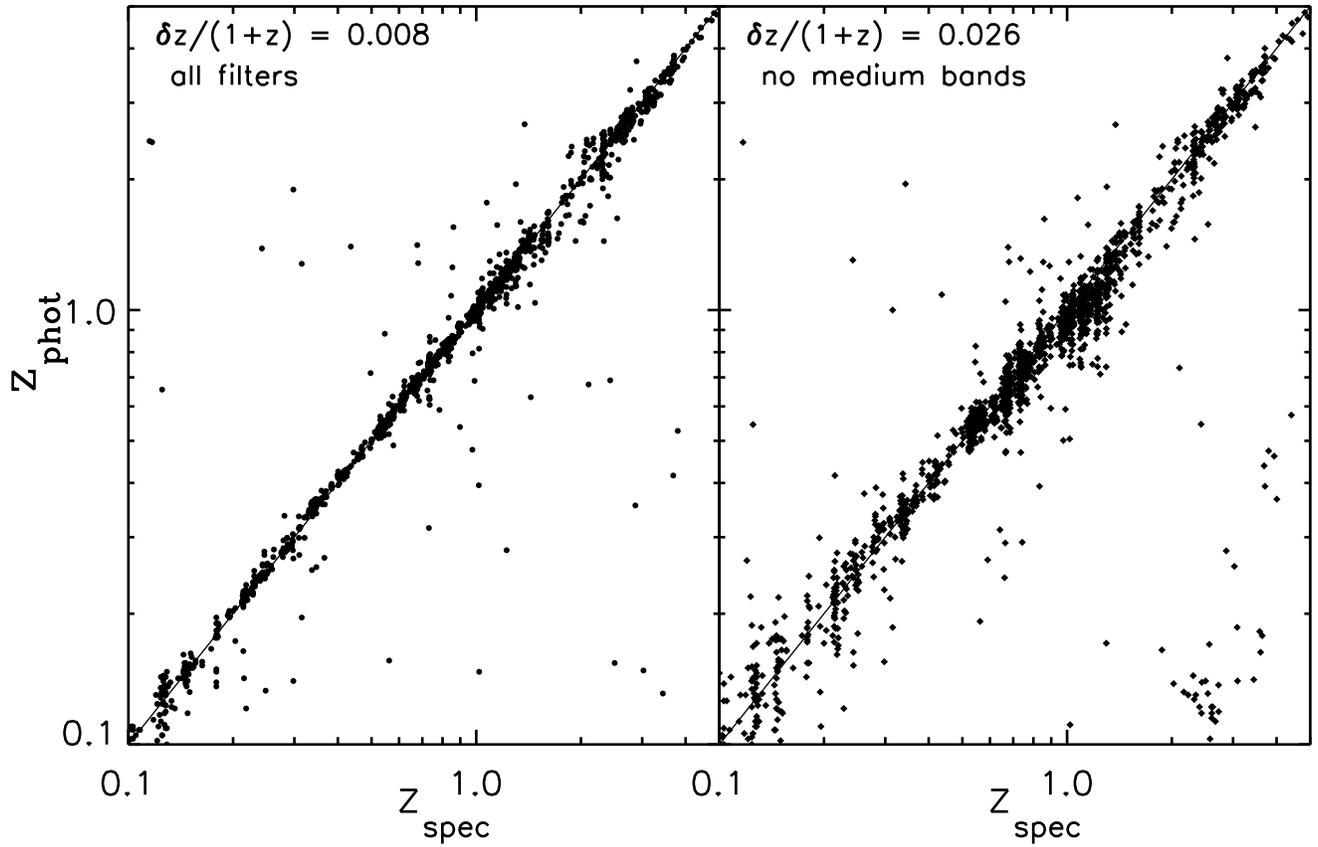}
\caption{Comparison of photometric redshift verses z\_spec for all sources with high quality spectroscopic redshifts and photometric redshift fits with $Q_z \le 1$.  Note that the left panel, which includes the additional medium band filters in the fit, lowers the overall dispersion between the spectroscopic and photometric redshifts.}
\label{eazycompnomed}
\end{figure*}

\begin{figure*}
\includegraphics[angle=0, width=0.99\textwidth]{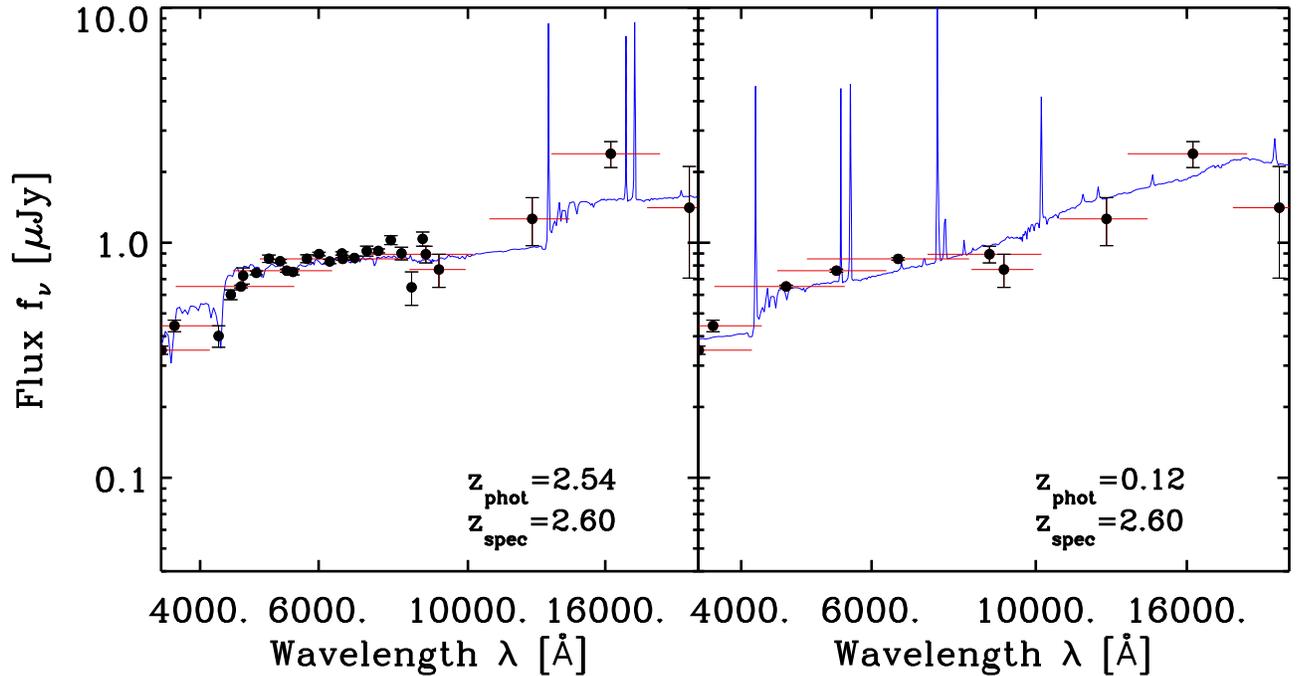}
\caption{An example source at z$\sim$2.5 and R$_{AB}\sim$24,  red lines  indicate the FWHM of the broad band filters.  Here the medium-band photometry increases the photometric redshift, ruling out the erroneous low-redshift solution found by fitting the SED with only broad-band photometry.  In the case of the broad-band photometry alone, a Balmer break is fit to the UBV bands; when the medium band photometry is added, this same region is found to fit the sharper  
Lyman alpha decrement instead. }
\label{eazycompnomedsed}
\end{figure*}

We quantify the effect of the medium-band filters, comparing the photometric redshifts with and without the additional information these filters provide.  
When excluding the medium-band filters, we retain the broad band optical filters, as well as the $J,H,K$ band and IRAC photometry.
Overall the medium-band filters provide a factor of three improvement over photometric redshifts using broad-band filters alone, decreasing the median $\Delta{\rm z/(1+z)}$ from 0.026 to 0.008 (comparing sources with $Q_z$ $\le$ 1).
We note that with the addition of the medium band filters, high quality ( $Q_z$ $\le$ 1) photometric redshifts are obtained for 20\% more of the spectroscopic redshift sample. 
The improvement is most noticeable at $0.1 \le z \le 1.2$ where the median$|{\Delta{\rm z/(1+z)}}|$ falls from 0.022 to 0.006 and at $z \ge 3.7$ where the median $\Delta{\rm z/(1+z)}$ falls from 0.024 to 0.006. 
These redshift intervals are where the Balmer break (3700 \AA) and Lyman limit (912 \AA) fall inside the region of the spectrum covered by the medium-band filters.
However, we still see improvement at $1.2 \le z \le 3.7$, where the median  $\Delta{\rm z/(1+z)}$ falls from 0.032 to 0.016.
In Figure \ref{eazycompnomed}, we compare the spectroscopic sources with good fits to the SEDs ($Q_z$ $\le$ 1) with (left) and without (right) the medium-band filters.
The medium-band filters not only tighten the accuracy around the z\_phot=z\_spec line, but also can help to rule out false redshift solutions (so called catastrophic failures)  for sources with z\_spec $\ge$ 1.2.  
In each of these cases at $2\sim 2.5$, the broad band photometry alone is fit with by a Balmer break feature in the SED at low redshift ($z\sim0.1$) but when the medium band photometry is added, the same optical region of the SED is fit by a Lyman alpha decrement caused by absorption by the IGM and the photometric redshift increases to a value more consistent with the spectroscopic determination.  
An example is shown in Figure \ref{eazycompnomedsed}, for a source with ${\rm R_{AB}\sim24}$ and ${\rm z_{spec}\sim2.5}$.

\subsection{Star / Galaxy Separation}
\label{sec:stars}
We use two methods to identify the stars in our catalog, the first using a $Bz'K$ color selection \citep{tayloretal2009} and the second fitting stellar SED templates \citep{Ilbertetal2009}.

\begin{figure}
\includegraphics[angle=0, width=0.47\textwidth]{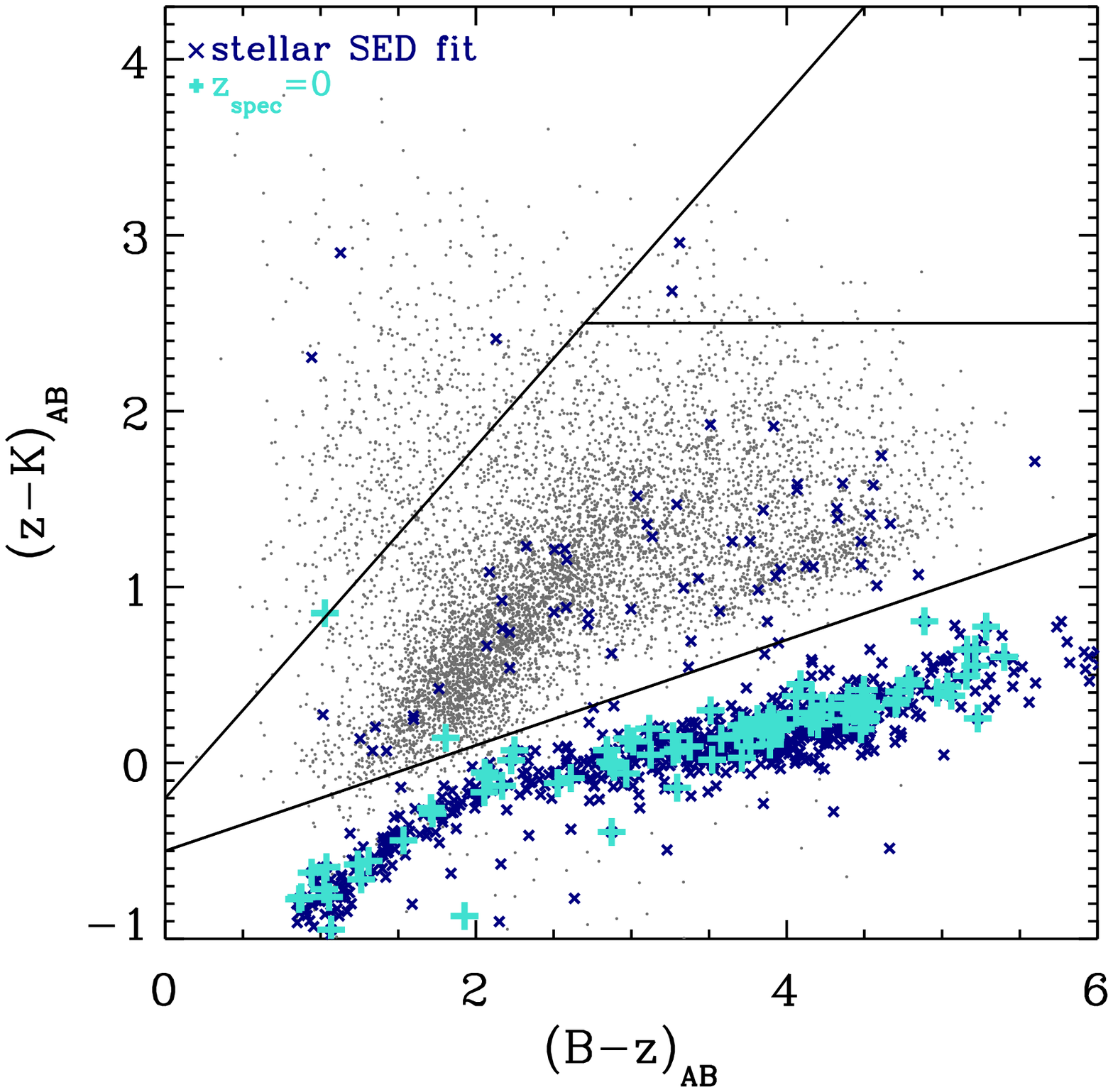}
\caption{Bzk color-color diagram indicating stars.  Note in this figure the z-K magnitude has been adjusted -0.04 and the B-z magnitude by + 0.56 to match the filter set used by \citet{daddietal2004}.  Sources with spectroscopic redshifts $z=0$ (turquoise  + ) and those which fit stellar SEDS (dark blue X), mostly lie along the stellar sequence in the bottom region on the figure. Also indicated are high redshift BzK  galaxies including star forming sBzK galaxies (top left) and passive pBzK galaxies (top right).  Overall, stars selected by template fitting to stellar SEDs are consistent with those indicated here on the BzK diagram.}
\label{fig:bzkstars}
\end{figure}

The $Bz'K$ diagram is traditionally used to select moderate redshift ($z\ga1.4$) galaxies \citep{daddietal2004}, but is also an efficient discriminating between stars and galaxies \citep[e.g.,][]{daddietal2004,blancetal2008,tayloretal2009}. 
Because our filter set is slightly different than that used by \citet{daddietal2004}, we apply the offsets determined by \citet{blancetal2008} of -0.04 mag in (z-K) and 0.56 mag in (B-z) to our colors before plotting them in the BzK diagram.
In Figure \ref{fig:bzkstars}, we show the $BzK$ diagram for at all sources in our catalog with $K_{AB} \le 21.84$ (i.e., $K_{vega} \le 20$).
The stellar sequence is clearly separated from the galaxies by the colors $z-K \le 0.3 (B-z)-0.5$, shown as a solid black line in Figure \ref{fig:bzkstars}.
For comparison, we include the handful of spectroscopically identified stars (large blue + in Figure 
\ref{fig:bzkstars}).
Only 2 of the 108 spectroscopically identified ($K_{AB} \le 21.84$) stars are missed by the $BzK$ star color selection.
Although this method of color selection is effective at identifying stars, we are limited to the sources bright enough to be detected in our relatively shallower K-band coverage.

To identify stars in all of our objects with photometric redshifts, we evaluate a $\chi^2$ value for $z=0$ stellar SED template  \citep{picklesetal1998} fits for each object.  Since the \citet{picklesetal1998} templates do not go past 2.5$\mu$m, we do not include the {\it Spitzer} IRAC bands in the stellar fits.  We flag as potential stars objects  with reduced $\chi^2_{star}\ge  \chi^2_{galaxy}$.   These sources are included in the $BzK$ diagram as blue crosses.  
The stars identified through template fitting agrees well with those selected in our $BzK$ color-selection, $\sim$90\% of stellar-template stars with  $K_{AB} \le 21.84$ have $BzK$ colors of stars (Figure \ref{fig:bzkstars}, dark blue crosses).
Additionally, 80\% of $BzK$-color-selected stars are also identified as stars by the stellar template fit.
Finally, we look at the subset of sources detected in the GEMS imaging \citep{haussieretal2007}.  
Selecting sources with stellar FWHM from the GEMS imaging, $\sim$80\% of them are selected by our template fitting method as stars, similar to the 84\% of point like sources that the COSMOS team identifies as stars through SED fitting \citep{Ilbertetal2009}.
Furthermore, looking at bright ($R\la24$) extended sources in the GEMS imaging, we find fewer that 2\% are misidentified as stars by this template fit classification.

\subsection{Comparison to Combo-17 Photometric Redshifts}
Combo-17 produced the first highly accurate photometric redshifts of the ECDF-S region, and we wish to compare our own photometric redshifts to theirs \citep{wolfetal2004}.
We restrict ourselves to the brighter photometry (r $\le$ 24) where their published photometric redshifts are more reliable (10\% accuracy), restricting ourseleves to r $\le$22, we have fewer sources for comparison, but the accuracy of Combo-17 (2\%) approaches our own \citep{wolfetal2004}.  
To examine our consistency, we compare our photometric redshifts to those determined in Combo-17 for all sources contained in both catalogs, finding $\Delta{\rm z/(1+z)} \sim 0.011$  at R $\le$ 22  and   $\Delta{\rm z/(1+z)} \sim 0.029$ at R $\le$24.
Therefore, our photometric redshift determinations are consistent with those determined by Combo-17 to within their errors.

\begin{figure}
\includegraphics[angle=0, width=0.47\textwidth]{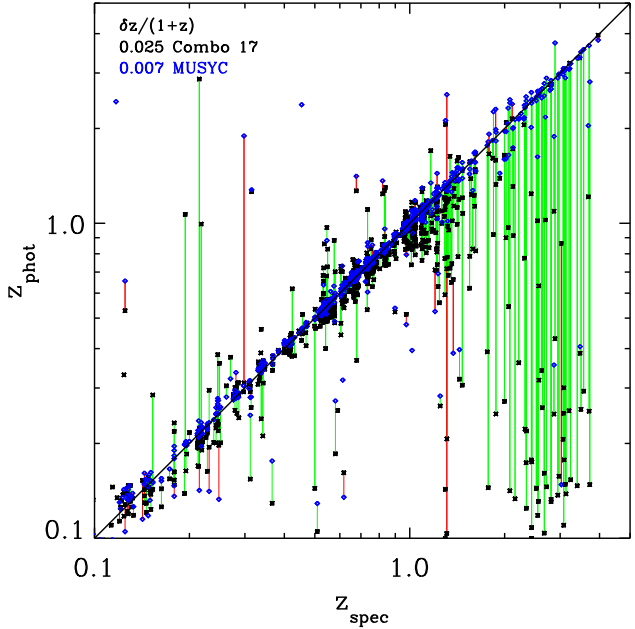}
\caption{Comparison of our photometric redshifts (blue diamonds) to those of Combo-17 (black X). Where our photometric redshift is closer to that determined from spectroscopy than that from Combo-17, a green line connects the two photometric redshift determinations.  Where the photometric redshift from Combo17 is closer to that determined from spectroscopy, a red line connects the two photometric redshift determinations.
Overall our photometric redshifts $\Delta{\rm z/(1+z)}$ are more accurate for many sources that were outliers in Combo-17. }
\label{fig:c17}
\end{figure}

To compare our respective accuracies, we use spectroscopic sources (Figure \ref{fig:c17}), and compare z\_spec with photometric redshifts for Combo-17 and our MUSYC survey. 
For  R $\le$ 22, we find  $\Delta{\rm z/(1+z)} \sim 0.011$ for Combo-17 and  $\Delta{\rm z/(1+z)} \sim 0.005$ for MUSYC, and considering all sources R$_{AB} \le$24 we find  $\Delta{\rm z/(1+z)} \sim 0.025$ for Combo-17 and  $\Delta{\rm z/(1+z)} \sim 0.007$ for our MUSYC redshifts. 
Therefore we find that although our photometric redshifts agree with those from Combo-17 survey to within their respective uncertainties, the larger number of deeper medium-band filter observations in the MUSYC survey improves the photometric redshift accuracy and extend this accuracy to fainter source populations.
We note that these differences come not only from the deeper medium-band data, but also by including JHK and IRAC photometry.


\section{The red sequence to $z\sim1$}
\label{sec:redseq}
Here we demonstrate the quality of our data by showing the color-magnitude
relation as a function of redshift in the E-CDFS.
The colors and morphologies of galaxies form a bimodal population \citep{blantonetal2003,baldryetal2004}.
This bimodality represents a fundamental relationship between mass and star formation history in galaxies, because the color of a galaxy is a proxy for the stellar age, where blue galaxies are younger and red galaxies are older.
The red sequence contains the brightest galaxies, which typically have early-type morphologies while the blue cloud galaxies typically have late-type morphologies.
Relatively
few galaxies are observed in the green valley, suggesting that galaxies spend only a short time here.
{\it How} the bimodal galaxy population was created is one of the outstanding questions in galaxy studies today.

Rest-frame fluxes were computed 
following the procedure of \citet{brammeretal2009}, which measures the rest-frame fluxes from the best-fitting template \citep{wolfetal2003}. 
This procedure is very
different from a $K$-correction as it uses the observed medium bands
that are closest in observed wavelength to the redshifted rest-frame
band of interest. The Brammer et al.\ (2009) methodology is embedded
in the EAZY photometric redshift code (Brammer et al.\ 2008).

\begin{figure*}
\includegraphics[angle=0, width=0.99\textwidth]{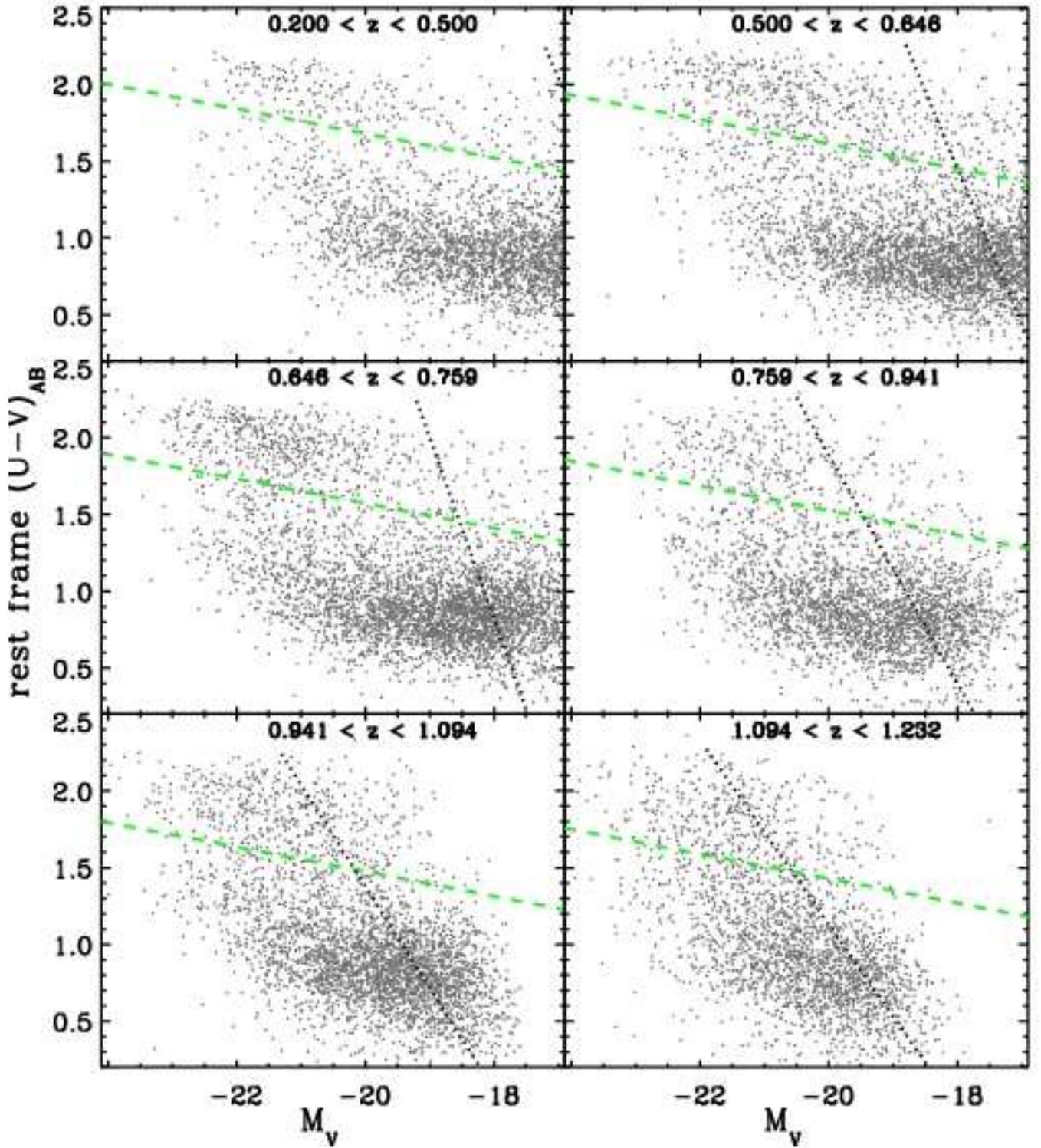}
\caption{Rest-frame color-magnitude diagrams
for galaxies as a function of redshift.  The redshift bins represent equal steps of co-moving volume, doubled in the last 3 bins.  The green line is the red sequence cut defined by \citet{belletal2004} and the dotted line shows the 90\% completeness limit.  The red sequence is clearly present out to z$\sim$1.2, where our redshift accuracy begins to fall.  The red sequence contains the most luminous galaxies, while the blue cloud dominates the overall number density, especially for the fainter galaxies.}
\label{fig:colormag}
\end{figure*}

In Figure \ref{fig:colormag}, we show the distribution of galaxies
in the rest-frame $U-V$ versus $V$ plane.
This plot most directly relates observables to the bi-modal color sequence as a function of redshift.
The restframe $U-V$ color covers the Balmer break and therefore is a good measure of recent star formation and the luminosity in the $V$ band is a proxy for stellar mass.
In Figure \ref{fig:colormag}, we also include the red sequence cut defined by \citet{belletal2004} in the Combo-17 survey (green dashed line) and our completeness limits in each redshift bin (blue solid line).
We see a  sequence of red points separated from the main blue clump of galaxies at all redshifts in our sample, highlighting the quality of our data.
The location of this red sequence is consistent with that found by \citet{belletal2004}.

\subsection{Passive Galaxy Evolution}

\begin{figure*}
\plotone{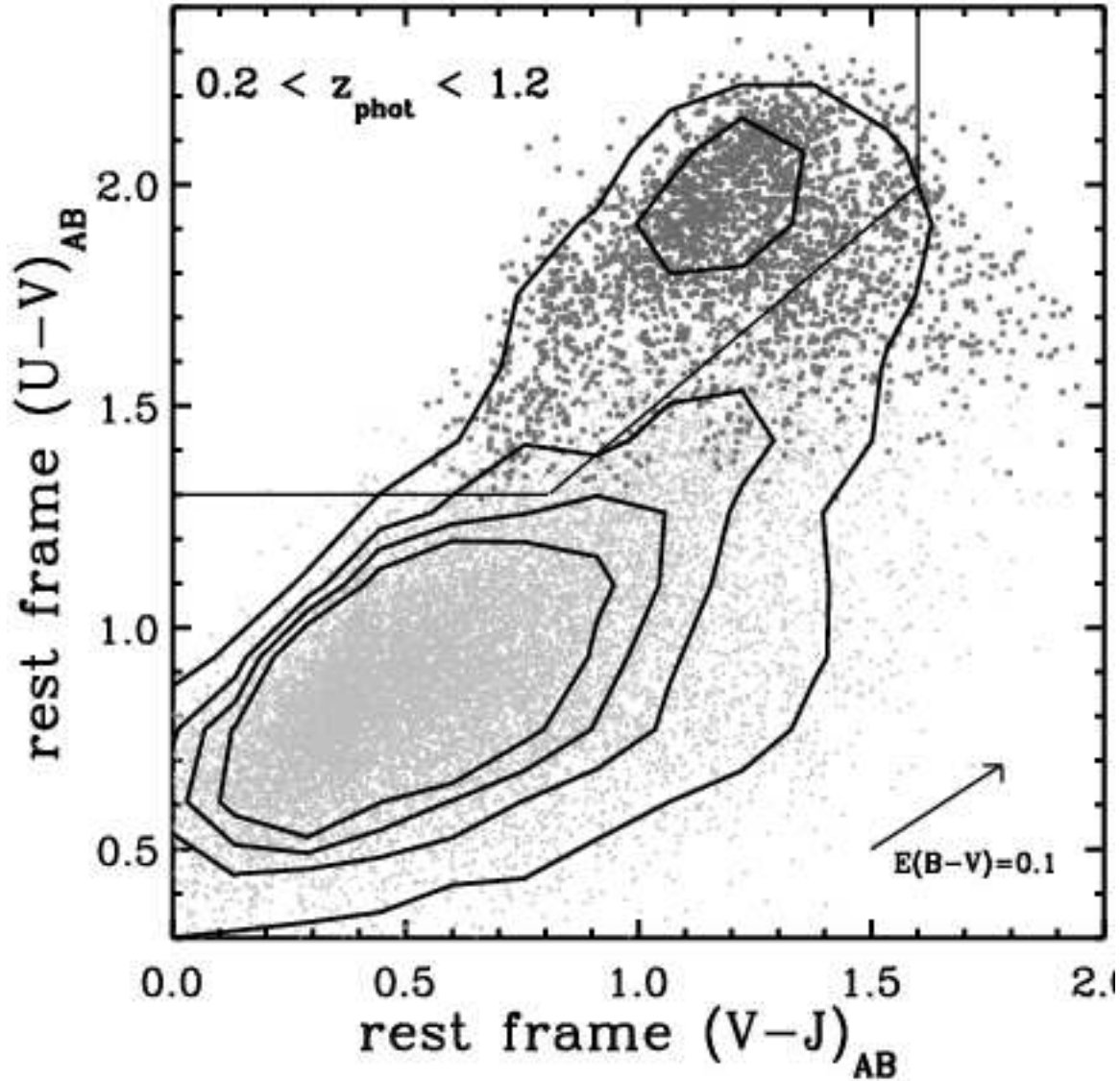}
\caption{Rest-frame color-color diagram, showing $U-V$ vs $V-J$ (Labbe
et al.\ 2005, Williams et al.\ 2009). A bimodal color-sequence is clearly visible separated by the dark lines \citep{williamsetal2009}. 
Dust-free quiescent galaxies are blue in $V-J$, therefore the peak of galaxies at red $U-V$ and bluer $V-J$ are passive red sequence galaxies. 
All galaxies with the \citet{belletal2004} red sequence cut are shown as darker points.
 This diagram identifies 20\% of those red sequence galaxies as dusty star forming galaxies rather than containing passive older stellar populations.}
\label{uvvj}
\end{figure*}

In color space both passive galaxies and dusty galaxies can appear red in $U-V$ colors.
To investigate the nature of the red sequence (Figure \ref{fig:colormag}), we use the IRAC photometry to separate out red passively evolving galaxies from red, dusty, star-forming galaxies.
The rest-frame $V-J$ color can distinguish between these two populations because dust-free galaxies are blue in V-J color, while the dust-obscured galaxies are still red \citep{labbeetal2005, wuytsetal2007}.
Following \citet{williamsetal2009}, we use a color-color diagram, rest-frame U-V vs V-J, to investigate the nature of the bimodal galaxy color sequence for all galaxies with 0.2 $\le$ z $\le$ 1.2 (Figure \ref{uvvj}).
We color-code the galaxies from the red sequence on the color-magnitude diagram red using \citet{belletal2004}'s red sequence cut.
As expected,
we find that passive and star forming galaxies
separate cleanly in color-color space \citep[see][]{williamsetal2009}.

We find that just over 20\% of the galaxies identified as red sequence
members in Figure \ref{fig:colormag} have rest-frame $V-J$ colors which
place them in the star-forming sequence, consistent with previous
work at these redshifts using morphological information \citep[e.g.,][]{ruhlandetal2009}
and with the results in \citep{brammeretal2009} and Whitaker et al.\
(2010, in prep). 
Splitting our redshift interval in two ($0.4<z<0.9$ and $0.9<z<1.2$) the fraction of red sequence galaxies with dusty colors is consistent between the two bins (21\% and 23\% respectively).

\section{Summary}
In this paper we present new deep 18-medium-band photometry in the well-studied E-CDFS field.
We reduced the raw data for the medium-band Subaru Suprime-Cam observations using a combination of standard routines and custom tasks.
The (public) catalog includes photometry from 10 ground-based broad-band images ($U$, $U38$, $B$, $V$, $R$, $I$, $z$, $J$, $H$, $K$), 4 IRAC images ($3.6$ $\mu$m, $4.5$ $\mu$m, $5.8$ $\mu$m, $8.0$ $\mu$m), and 18 medium-band images ($IA427$, $IA445$, $IA464$, $IA484$, $IA505$, $IA527$, $IA550$, $IA574$, $IA598$, $IA624$, $IA651$, $IA679$, $IA709$, $IA738$, $IA767$, $IA797$, $IA856$). 
The full catalog provides multiwavelength SEDs for $\sim$80,000 galaxies in the ECDF-S down to $R_{[AB]}\sim27$ (40,000 at R$\le$25.2, the median depth of the medium band imaging), although the accuracy and completeness of the photometric redshifts declines at $R\gtrsim 25.5$.

We computed accurate photometric redshifts using EaZy, a public photometric redshift code.
The addition of the medium band proves a factor of four improvement in the photometric redshift accuracy over the use of broad band filters alone.
Comparing to spectroscopic redshifts, we find a scatter in $\Delta{\rm z/(1+z)}$
of 0.008 for the full sample, $0.006$ at $0.2<z<1.2$ where the Balmer break is covered by the medium band filters and 0.01 at $z\ge3.7$ where the Lyman Limit (912\AA) is covered by the medium band filters.
We find that the additional filters in the optical region even improve the photometric redshifts for sources at $1.2\ge z\ge3.7$, decreasing the median $\Delta{\rm z/(1+z)}$ from 0.03 to 0.02.

We demonstrate that these photometric redshifts are sufficiently accurate to determine precise rest-frame colors of the galaxies.
We detect the bimodal galaxy distribution out to $z\sim1.2$ and find that 20\% of the galaxies on the red-sequence have longer wavelength colors consistent with being dusty. In keeping with the spirit of the MUSYC collaboration,
we provide public access to the images, photometry, and photometric redshifts.
In future papers we will combine these accurate redshifts with the extensive ancillary data in the ECDF-S.

\section{Acknowledgements}
We thank the referee for helpful comments which improved this paper.
Support from NSF grants AST-0407295, AST-0449678, AST-0807570, 
NASA JPL grants 1277255 and 1282692,
and Yale University is gratefully acknowledged.
Support for the work of KS was provided by NASA through Einstein Postdoctoral
Fellowship grant number PF9-00069 issued by the Chandra X-ray Observatory
Center, which is operated by the Smithsonian Astrophysical Observatory for and
on behalf of NASA under contract NAS8-03060.
We also thank Shunju Sasaki for help with the imaging and data reductions.


\end{document}